\documentclass{ws-ijmpa}

\begin{document}

\markboth{M.-C. Chen, S. Dawson \& T. Krupovnickas}
{Constraining New Models with Precision Electroweak Data}

\catchline{}{}{}{}{}

\title{Constraining New Models with Precision Electroweak Data
}

\author{\footnotesize Mu-Chun Chen}

\address{Theoretical Physics Department, 
Fermi National Accelerator Laboratory\\
Batavia, IL 60510, USA\\
mcchen@fnal.gov}

\author{Sally Dawson and Tadas Krupovnickas}

\address{High Energy Theory Group, Brookhaven National Laboratory\\ 
        Upton, New York 11973, U.S.A.\\
dawson@quark.phy.bnl.gov, tadas@quark.phy.bnl.gov
}

\maketitle

\pub{Received (Day Month Year)}{Revised (Day Month Year)}

\begin{abstract}
Electroweak precision data have been extensively 
used to constrain models containing physics beyond that 
of the Standard Model. When the model contains Higgs 
scalars in representations other than singlets or 
doublets, and hence $\rho\ne 1$ at tree level, a correct 
renormalization scheme requires more inputs than the 
three commonly used for the Standard Model case.  
In such cases, the one loop electroweak results cannot be 
split into a Standard Model contribution plus a piece which vanishes 
as the scale of new physics becomes much larger than $M_W$.  We 
illustrate our results by presenting the dependence of $M_W$ on the 
top quark mass in a model with a Higgs triplet and in the 
$SU(2)_L \times SU(2)_R$ left-right symmetric
model. In these models, the allowed range for the lightest neutral  
Higgs mass can be as large as a few TeV.
\keywords{Beyond Standard Model, Higgs Physics}
\end{abstract}

\ccode{PACS Nos.: 14.80.Cp, 12.15.Lk}

\section{Introduction}

Measurements at LEP, SLD, and the Tevatron have been used extensively to limit
models with physics beyond that of the Standard Model (SM)\cite{ewpewwg}. 
By performing global fits to a series of precision measurements, 
information about the parameters of new models can be 
inferred\cite{Peskin:1991sw,altarelli:1991a}. 
The simplest example of this approach is the prediction of the $W$ boson mass. 
In the Standard Model, 
the $W$- boson mass, $M_W$, can be predicted in terms of other parameters
of the theory.  The predicted $W$ boson mass is
strongly correlated with the experimentally measured value of
  the top quark mass, $m_t$,  
and increases quadratically as the top quark mass is increased.  
This strong correlation between $M_W$ and $m_t$ in the Standard Model
can be used to limit the allowed region for the Higgs 
boson mass\cite{pdg04}.

In a model with Higgs particles in
representations other than $SU(2)$ doublets and singlets, there
are more parameters in the gauge/Higgs sector than in the Standard 
Model. The SM tree level relation, $\rho=M_W^2/(M_Z^2 c_\theta^2)=1$ 
no longer holds and when the theory is renormalized at one loop,
models of this type  will require extra 
input parameters\cite{Passarino:1990xx,Lynn:1990zk,chivukula:1999}.
Models with new physics are often written in terms of the SM Lagrangian,
$L_{SM}$ plus an extra contribution,
\begin{equation}
L=L_{SM}+L_{NP}
\end{equation}
where $L_{NP}$ represents contributions from new physics beyond
the SM.  Phenomenological
studies  have then considered the contributions of $L_{SM}$ at one-loop,
plus the tree level contributions of $L_{NP}$.  In this note, we 
give two specific examples with  $\rho\ne 1$ at tree
level, where we demonstrate that this  procedure is incorrect. 
We discuss in detail what happens in these models when the scale
of the new physics becomes much larger than the electroweak
scale and demonstrate explicitly that the SM is not recovered.

The possibility of a heavy Higgs boson which is consistent with precision
electroweak data has been considered by Chivukula, Hoelbling and Evans\cite{Chivukula:2000px} 
and by Peskin and Wells\cite{Peskin:2001rw} 
in the context of oblique corrections.  In terms
of the $S$, $T$ and $U$ parameters\cite{Peskin:1991sw,altarelli:1991a},  
a large contribution to isospin violation,  $\delta \rho =\alpha T>1$, 
can offset the contribution of a heavy Higgs boson  to electroweak
observables such as the $W$ boson mass. The triplet model considered in
this paper provides an explicit realization of this mechanism.
The oblique parameter formulation neglects contributions to observables from
vertex and box diagrams, which are numerically important in the example
discussed here.

In Section~\ref{renorm}, we review the important features of the SM
for our analysis.
We discuss two examples in Sections~\ref{higgstrip} 
and Appendix~\ref{lrmodel} where the new physics does
not decouple from the SM at one-loop.  
For simplicity, we consider 
only the dependence of the $W$ boson mass on the top quark mass 
and demonstrate that a correct renormalization scheme gives 
very different results from the SM result in these models. 
Section~\ref{higgstrip} contains 
a discussion of the SM augmented by a real scalar triplet, 
and Appendix~\ref{lrmodel} contains a discussion of a 
left-right $SU(2)_{L}\times SU(2)_{R}$ symmetric model.  
In Section~\ref{nondecoupling}, we show that the dependence on scalar masses  
in the W-boson mass is quadratic and demonstrate that 
the triplet is non-decoupling. Our major results are summarized in 
Eq.~\ref{cc1}-\ref{cc3}.   
These results are novel and have not been discussed in the literature before. 
Section~\ref{results} contains our numerical 
results and Section~\ref{conc}
concludes this paper.  Similar results in the context of the littlest
Higgs model have previously been found in Ref.~\refcite{Chen:2003fm,Chen:2004ig}.

\section{Renormalization}
\label{renorm}

The one-loop renormalization of the SM has been 
extensively studied\cite{Marciano:1980pb,marciano:1980a,Sirlin:1981yz} and we
present only a brief summary here, in order to set the stage 
for Sections~\ref{higgstrip} and Appendix \ref{lrmodel}.  
In the electroweak sector of the SM, the 
gauge sector has three fundamental 
parameters, the $SU(2)_{L} \times U(1)_{Y}$ gauge coupling constants, 
$g$ and $g^\prime$, as well as the vacuum expectation (VEV) of the 
Higgs boson,  $v$. Once these three parameters are fixed, 
all other physical quantities in the gauge sector can be derived in terms 
of these three parameters and their counter terms\footnote{There
are of course also the fermion masses and the Higgs boson mass.  The
renormalization of these quantities does not affect our discussion. We assume
that the contributions from Higgs tadpole graphs can be set to zero with an
appropriate renormalization condition.}. 
We can equivalently choose the muon decay constant, $G_{\mu}$, 
the Z-boson mass, $M_{Z}$, and the fine structure constant 
evaluated at zero momentum, $\alpha \equiv \alpha(0)$, 
as our input parameters. 
Experimentally, the measured values for these input 
parameters are\cite{pdg04},
\begin{eqnarray}
G_{\mu} & = & 1.16637(1) \times 10^{-5} \; \mbox{GeV}^{-2}\\
M_{Z} & = & 91.1876(21) \; \mbox{GeV}\\
\alpha & = & 1/137.036 \quad .
\end{eqnarray}

The W-boson mass then can be defined through muon 
decay\cite{Marciano:1980pb,jegerlehner}, 
\begin{equation}
G_{\mu} = \frac{\pi \alpha}{\sqrt{2} M_{W}^{2} s_{\theta}^{2}}
\biggl[1 + \Delta r_{SM} \biggr]
\end{equation}
where $\Delta r_{SM}$ summarizes the radiative corrections,
\begin{equation}
\Delta r_{SM} = - \frac{\delta G_{\mu}}{G_{\mu}} 
+ \frac{\delta \alpha}{\alpha}
- \frac{\delta s_{\theta}^{2}}{s_{\theta}^{2}} 
- \frac{\delta M_{W}^{2}}{M_{W}^{2}}\quad , 
\label{drdef}
\end{equation}
where $s_{\theta}=\sin\theta_{W}$, $c_{\theta}=\cos\theta_{W}$ and 
$\theta_{W}$ is the weak mixing angle. 
The SM satisfies $\rho =1$ at tree level,
\begin{equation}
\rho = 1 =\frac{M_{W}^{2}}{M_{Z}^{2}\overline{c}_{\theta}^{2}}\quad .
\label{rhodef}
\end{equation}
In Eq.~(\ref{rhodef}), $M_W$ and $M_Z$ are the physical gauge boson 
masses, and so our definition of the weak mixing angle, $\overline{s}_\theta$, 
corresponds to the on-shell scheme\cite{Erler:2004nh}. 
It is important to note that
in the SM, $\overline{s}_\theta$ is not a free parameter, but is derived from
\begin{equation}\label{swos}
\overline{s}_\theta^2=1-{M_W^2\over M_Z^2}\quad .
\end{equation}

The counterterms of Eq.~(\ref{drdef}) are given by\cite{jegerlehner,pierce},
\begin{eqnarray}
\frac{\delta G_{\mu}}{G_{\mu}} 
& = & - \frac{\Pi_{WW}(0)}{M_{W}^{2}}+\delta_{V-B}
\\
\frac{\delta \alpha}{\alpha}
& = & \Pi_{\gamma\gamma}^{\prime}(0) 
+ 2 \frac{\overline{s}_{\theta}}{\overline{c}_{\theta}} 
\frac{\Pi_{\gamma Z}(0)}{M_{Z}^{2}}
\\
\frac{\delta M_{W}^{2}}{M_{W}^{2}}
& = & \frac{\Pi_{WW}(M_{W}^{2})}{M_{W}^{2}}\quad ,
\end{eqnarray}
where $\Pi_{XY}$, for $(XY=WW, ZZ, \gamma \gamma, \gamma Z)$, 
are the gauge boson 2-point functions;   
$\Pi^{\prime}_{\gamma \gamma} (0)$ is defined as   
$\frac{d\Pi_{\gamma \gamma}(p^{2})}{dp^{2}}\bigg|_{p^{2}=0}$. 
The term $\delta_{V-B}$ contains the box and vertex
contributions to the renormalization of $G_\mu$\cite{jegerlehner,pierce}. 

The counterterm for $\overline{s}_{\theta}^{2}$ 
can be derived from Eq.~(\ref{rhodef}), 
\begin{equation}
\frac{\delta \overline{s}_{\theta}^{2}}{\overline{s}_{\theta}^{2}}
=  \frac{\overline{c}_{\theta}^{2}}{\overline{s}_{\theta}^{2}} 
\biggl[ \frac{\delta M_{Z}^{2}}{M_{Z}^{2}} 
- \frac{\delta M_{W}^{2}}{M_{W}^{2}}
\biggr]
= \frac{\overline{c}_{\theta}^{2}}{\overline{s}_{\theta}^{2}}
\biggl[ 
\frac{\Pi_{ZZ}(M_{Z}^{2})}{M_{Z}^{2}} 
- \frac{\Pi_{WW}(M_{W}^{2})}{M_{W}^{2}}
\biggr]\quad .
\label{stdef}
\end{equation}
Putting these contributions together we obtain,
\begin{eqnarray}\label{drsm1}
\Delta r_{SM} & = & \frac{\Pi_{WW}(0)-\Pi_{WW}(M_{W}^{2})}{M_{W}^{2}}
+ \Pi_{\gamma\gamma}^{\prime}(0) 
+ 2 \frac{\overline{s}_{\theta}}{\overline{c}_{\theta}} 
\frac{\Pi_{\gamma Z}(0)}{M_{Z}^{2}}
\\
&&
- \frac{\overline{c}_{\theta}^{2}}{\overline{s}_{\theta}^{2}}
\biggl[
\frac{\Pi_{ZZ}(M_{Z}^{2})}{M_{Z}^{2}} - \frac{\Pi_{WW}(M_{W}^{2})}{M_{W}^{2}}
\biggr]\quad .\nonumber
\end{eqnarray}
These gauge boson self-energies can be found in 
Ref.~\refcite{Chen:2003fm} and \refcite{Hollik:1993cg,degrassi:1992a} and we note that the fermion and scalar 
contributions to the two-point function $\Pi_{\gamma Z}(0)$ vanish.  
The dominant contributions to $\Delta r_{SM}$ is from the top quark, 
and the contributions of the top and bottom quarks to the gauge boson 
self-energies are given in \ref{loop}.
The $m_{t}^{2}$ dependence in $\Pi_{WW}(0)$ and $\Pi_{WW}(M_{W}^{2})$ 
exactly cancel, thus the difference, $\Pi_{WW}(0)-\Pi_{WW}(M_{W}^{2})$, 
depends on $m_{t}$ only logarithmically. 
The second term, $\Pi_{\gamma \gamma}^{\prime}(0)$, also depends on $m_{t}$ 
logarithmically.   However, the quadratic $m_{t}^{2}$ dependence 
in $\frac{\Pi_{ZZ}(M_{Z}^{2})}{M_{Z}^{2}} $ and 
$\frac{\Pi_{WW}(M_{W}^{2})}{M_{W}^{2}}$ do not cancel. 
Thus $\Delta r_{SM}$ depends on $m_{t}$ quadratically, and is given by the
well known result, keeping only the two-point functions 
that contain a quadratic dependence on $m_{t}$\cite{Chanowitz:1978uj,chanowitz:1978a},
\begin{eqnarray}
\Delta r_{SM}^t& \simeq &-\frac{\delta G_{\mu}}{G_{\mu}} 
- \frac{\delta M_{Z}^{2}}{M_{Z}^{2}}
- \biggl( \frac{\overline{c}_{\theta}^{2}
-\overline{s}_{\theta}^{2}}{\overline{c}_{\theta}^{2}} \biggr)
\frac{\delta \overline{s}_{\theta}^{2}}{\overline{s}_{\theta}^{2}}\nonumber \\
& \simeq & - {G_\mu\over \sqrt{2}}{N_c\over 8 \pi^2}
\biggl( {\overline{c}_\theta^2\over \overline{s}_\theta^2}
\biggr) m_t^2\quad ,
\end{eqnarray}
where $N_c=3$ is the number of colors and the superscript $t$
denotes that we have included only the top quark contributions, 
in which the dominant contribution is quadratic.
The complete contribution to $\Delta r_{SM}$ can be
approximated,
\begin{equation}
\Delta r_{SM} \simeq .067 + \Delta r_{SM}^t + {\alpha\over \pi
\overline{s}_\theta^2}{11\over 48} \biggl( 
\ln \biggl({M_H^2\over M_Z^2}\biggr)
- \frac{5}{6} \biggr)
+{\hbox{2-loop contributions}}\quad .
\label{drsm}
\end{equation}
The first term in Eq.~(\ref{drsm}) results from the scaling of 
$\delta \alpha$ from zero momentum to $M_{Z}$\cite{Marciano:2004hb}. 
In the numerical results, the complete contributions to 
$\Delta r_{SM}$ from top and bottom quarks, 
the Higgs boson as well as the gauge bosons are included, 
as given in Eq.~(\ref{drsm1}).

\section{Standard Model with an additional 
$SU(2)_{L}$ triplet Higgs boson}
\label{higgstrip}

In this section, we consider the SM with an additional Higgs
 boson transforming
as a real triplet (Y=0) under the $SU(2)_L$ gauge symmetry. 
Hereafter we will call this the Triplet Model (TM)\cite{Logan:1999if,logan:2000a}. 
This model has been 
considered at one-loop by Blank and Hollik\cite{Blank:1997qa} 
and we have checked that our numerical codes are correct by reproducing 
their results. 
In addition, we derive the scalar mass dependence in this model and 
show that the triplet is non-decoupling by investigating various 
scalar mass limits. We also find the conditions under which 
the lightest neutral Higgs can be as heavy as a TeV, which has 
new important implications on Higgs searches. These results concerning 
the scalar fields 
are presented in the next section.

The $SU(2)_{L}$ Higgs doublet 
in terms of its component fields is given by,
\begin{equation}
H = \left(\begin{array}{c}
\phi^{+}\\
\frac{1}{\sqrt{2}}(v + \phi^{0} + i \phi_{I}^{0})
\end{array}\right) \; ,
\end{equation}
with $\phi^{0}$ being the Goldstone boson corresponding 
to the longitudinal component of the $Z$ gauge boson. 
A real $SU(2)_L$ triplet, $\Phi$, can be written
as $(\eta^+,\eta^0, \eta^-)$, 
\begin{equation}
\Phi = \left(\begin{array}{c}
\eta^{+}\\
v^{\prime} + \eta^{0}\\
-\eta^{-}
\end{array}\right) \; .
\end{equation}
There are thus four physical Higgs fields in the spectrum: 
There are two neutral Higgs bosons, 
$H^{0}$ and $K^{0}$, 
\begin{eqnarray}\label{neutralmix}
\left(\begin{array}{c}
H^{0}\\ K^{0}
\end{array}\right)
& = &
\left(\begin{array}{cc}
c_{\gamma} & s_{\gamma}\\
-s_{\gamma} & c_{\gamma}
\end{array}\right) 
\left(\begin{array}{c}
\phi^{0}\\ \eta^{0}
\end{array}\right)\; ,
\end{eqnarray}
and the mixing between the two neutral Higgses is described 
by the angle $\gamma$. The charged Higgses $H^{\pm}$ are  
linear combinations of the charged components in 
the doublet and the triplet, with a mixing angle $\delta$, 
\begin{eqnarray}\label{chargedmix}
\left(\begin{array}{c}
G^{\pm}\\ H^{\pm}
\end{array}\right)
& = &
\left(\begin{array}{cc}
c_{\delta} & s_{\delta}\\
-s_{\delta} & c_{\delta}
\end{array}\right) 
\left(\begin{array}{c}
\phi^{\pm}\\ \eta^{\pm}
\end{array}\right)\; ,
\end{eqnarray}
where $G^{\pm}$ are the Goldstone bosons corresponding 
to the longitudinal components of $W^{\pm}$. 
The masses of these four physical scalar fields, 
$M_{H^{0}}$, $M_{K^{0}}$ and $M_{H^{\pm}}$, respectively, 
are free parameters in the model. 
The $W$ boson mass is  given by, 
\begin{equation}
M_W^{2}={g^2\over 4}(v^2+v^{\prime2}) \; ,
\end{equation}
where $v/\sqrt{2}=\langle \phi^0\rangle $ 
is the VEV of the neutral component of the $SU(2)_L$ 
Higgs boson and $v^{\prime} = \langle \eta^0 \rangle = \frac{1}{2}
v \tan\delta~$ is the vacuum expectation value of the additional 
scalar, leading to the relationship
$v_{SM}^2=(246~\mbox{GeV})^2=v^2+v^{\prime 2}$.  A real triplet 
does not contribute to $M_Z$, leading to 
\begin{equation}
\rho 
= 1+4 {v^{\prime 2}\over v^2} = \frac{1}{c_{\delta}^{2}} \; .
\end{equation}
The main result of this section is to show that the renormalization of 
a theory with $\rho \ne 1$ at tree level is fundamentally different
from that of the SM.
  
Due to the presence of the $SU(2)_{L}$ triplet Higgs, the gauge sector now 
has four fundamental parameters, the additional parameter being the VEV of 
the $SU(2)_{L}$ triplet Higgs, $v^\prime$. A consistent renormalization 
scheme thus requires a fourth input parameter\cite{Blank:1997qa}.
We choose the fourth input parameter to be the effective leptonic 
mixing angle, $\hat{s}_{\theta}$, 
which is defined as the ratio of the vector to axial vector 
parts of the $Ze\overline{e}$ coupling,
\begin{equation}
L=-i {\overline e} (v_e +\gamma_5 a_e)\gamma_\mu e Z^\mu\quad ,
\end{equation}
with $v_e={1\over 2}-2 \hat{s}_{\theta}^{2}$ 
and $a_e={1\over 2}$.  This leads
to the definition of $\hat{s}_{\theta}$,
\begin{equation}\label{swdef}
1-4\hat{s}_{\theta}^{2} 
= \frac{\mbox{Re}(v_e)}{\mbox{Re}(a_{e})} \; .
\end{equation}
The measured value from LEP is given by 
$\hat{s}_{\theta}^{2} = 0.23150 \pm 0.00016$\cite{ewpewwg}.

As usual the $W$ boson mass is defined through muon 
decay\cite{Marciano:1980pb,jegerlehner},  
\begin{equation}
G_{\mu} = \frac{\pi \alpha(M_{Z})}{\sqrt{2} M_{Z}^{2} 
\hat{c}_{\theta}^2 
\hat{s}_{\theta}^{2}
\rho \bigl(1 - \Delta r_{triplet} \bigr)}
 \; ,
\end{equation}
where we have chosen $\alpha(M_Z)$ instead of $\alpha(0)$ 
as an input parameter. 
The contribution to $\Delta r_{triplet}$ is similar to that
of the SM,
\begin{equation} 
\Delta r_{triplet} = -\frac{\delta G_{\mu}}{G_{\mu}} 
- \frac{\delta M_{Z}^{2}}{M_{Z}^{2}}
+ \frac{\delta \alpha}{\alpha(M_{Z})}
- \biggl( \frac{\hat{c}_{\theta}^{2}
-\hat{s}_{\theta}^{2}}
{\hat{c}_{\theta}^{2}} \biggr)
\frac{\delta \hat{s}_{\theta}^{2}}
{\hat{s}_{\theta}^{2}}
-{\delta\rho\over \rho} \; ,
\end{equation}
where the counter term $\delta \rho$ is defined through $M_{W}$, 
$M_{Z}$ and $\hat{s}_{\theta}$ as, 
\begin{equation}
{\delta \rho\over\rho}
={\delta M_W^2\over M_W^2}-{\delta M_Z^2\over
M_Z^2}+\biggl({\hat{s}_{\theta}^{2}\over \hat{c}_{\theta}^{2}}\biggr)
{\delta \hat{s}_{\theta}^{2} \over \hat{s}_{\theta}^{2}} \; .
\end{equation}
Unlike in the SM case where $\overline{s}_{\theta}$ is defined through 
$M_{W}$ and $M_{Z}$ as given in Eq.~(\ref{stdef}),  
now $\hat{s}_{\theta}$ is an independent parameter, and its 
counter term is given by\cite{Blank:1997qa},
\begin{eqnarray}
\frac{\delta \hat{s}_{\theta}^{2}}{\hat{s}_{\theta}^{2}} & = & 
Re \biggl[ \; \biggl(\frac{\hat{c}_{\theta}}{\hat{s}_{\theta}} 
\biggr)\; \biggr[
\frac{\Pi_{\gamma Z}(M_{Z}^{2})}{M_{Z}^{2}}
+ \frac{v_{e}^{2}-a_{e}^{2}}{a_{e}} \Sigma_{A}^{e}(m_{e}^{2}) 
\nonumber\\
& & 
- \frac{v_{e}}{2\hat{s}_{\theta}\hat{c}_{\theta}}
\biggl( \frac{\Lambda_{V}^{Z\overline{e}{e}}(M_{Z}^{2})}{v_{e}}
-\frac{\Lambda_{A}^{Z\overline{e}{e}}(M_{Z}^{2})}{a_{e}} \biggr) \;
\biggr]\;
\biggr]  \; , \nonumber \\
& \equiv & \biggl(\frac{\hat{c}_{\theta}}{\hat{s}_{\theta}} \biggr)
\frac{Re\biggl(\Pi^{\gamma Z}(M_{Z}^{2})\biggr)}{M_{Z}^{2}} 
+ \delta_{V-B}^{\prime} \; ,
\end{eqnarray}
where $\Sigma_{A}^{e}$ is the axial part of the electron self-energy, 
$\Lambda_{V}^{Z\overline{e}{e}}$ and $\Lambda_{A}^{Z\overline{e}{e}}$ 
are the vector and axial-vector form factors of 
the vertex corrections to the $Z\overline{e}e$ coupling. These effects 
have been included in our numerical 
results\cite{Blank:1997qa,Consoli:1989pc}.  
The total correction to $\Delta r_{triplet}$ in this case is then 
given by,
\begin{eqnarray}
\Delta r_{triplet} & = & \frac{\Pi_{WW}(0) - \Pi_{WW}(M_{W}^{2})}{M_{W}^{2}}
+ \Pi_{\gamma\gamma}^{\prime}(0) 
+ 2 \frac{\hat{s}_{\theta}}{\hat{c}_{\theta}}
\frac{\Pi_{\gamma Z}(0)}{M_{Z}^{2}}
\\
&&
- \frac{\hat{c}_{\theta}}{\hat{s}_{\theta}} 
\frac{\Pi_{\gamma Z}(M_{Z}^{2})}{M_{Z}^{2}}
+ \delta_{V-B} + \delta_{V-B}^{\prime}
\; , \nonumber
\end{eqnarray}
where $\delta_{V-B}$ summarizes the vertex and box corrections 
in the TM model, 
and it is given by\cite{Blank:1997qa}, 
\begin{equation}
\delta_{V-B} = \frac{\alpha}{4\pi \hat{s}_{\theta}^{2}}
\biggl[ 6 + 
\frac{10-10\hat{s}_{\theta}^{2}-3(R/\hat{c}_{\theta}^{2})
(1-2\hat{s}_{\theta}^{2})}{2(1-R)}
\ln R \biggr], \qquad 
R \equiv M_{W}^{2} / M_{Z}^{2} \; ,
\end{equation}
where we show only the finite contributions in the above equation.
Keeping only the top quark contribution,
\begin{eqnarray}
{\delta \hat{s}_{\theta}^{2} \over \hat{s}_{\theta}^{2}}=
 \biggl({\hat{c}_{\theta} \over \hat{s}_{\theta}}\biggr)
\frac{\Pi_{\gamma Z}(M_{Z}^{2})}{M_{Z}^{2}}
=
-{\alpha\over \pi \hat{s}_{\theta}^{2}}
\biggl({1\over 2} -{4\over 3} \hat{s}_{\theta}^{2}
\biggr)\biggl\{{1\over 3}\biggl( \ln {Q^2\over m_t^2} +{1\over\epsilon}
\biggr)-2 I_3\biggl({M_Z^2\over m_t^2}\biggr)\biggr\} \; ,
\end{eqnarray}
where $Q$ is the momentum cutoff in dimensional regularization and 
the definition of the function $I_{3}$ can be found in Appendix A. 
As $\Pi_{\gamma Z}(M_{Z}^{2})$ is logarithmic, the $m_{t}$ 
dependence of $M_{W}$ is now logarithmic. 
We note that this much softer relation between $M_{W}$ and $m_{t}$ 
is independent of the choice of the fourth input parameter. 
This will become clear in our second example, the left-right symmetric model, 
given in \ref{lrmodel}. 
In our numerical results, we have included in 
$\Delta r_{triplet}$ 
the complete contributions, which are summarized 
in \ref{scalar}, from the top and bottom quarks and  
the four scalar fields, as well as the gauge bosons, 
and the complete set of vertex and box corrections. 

\section{Non-decoupling of the Triplet}
\label{nondecoupling}

As shown in \ref{scalar}, $\Delta r_{triplet}$ 
depends on scalar masses quadratically. This has important implications 
for models with triplets, such as the littlest Higgs 
model\cite{Chen:2003fm,Chen:2004ig}. 
The two point function 
$\Pi_{\gamma Z}(0)$ does not have any scalar dependence, while  
$\Pi_{\gamma\gamma}^{\prime}(0)$ and $\Pi_{\gamma Z}(M_{Z})$ depend on 
scalar masses logarithmically. The quadratic dependence thus comes 
solely from the function $\Pi_{WW}(0)-\Pi_{WW}(M_{W})$. 
When there is a large hierarchy among  
the three scalar masses (case (c) in 
\ref{scalar} and its generalization), 
all contributions are of the same sign, and are 
proportional to the scalar mass squared, 
\begin{eqnarray}\label{cc1}
\Delta r_{triplet}^{S} &\rightarrow&
\frac{\alpha}{4\pi \hat{s}_{\theta}^{2}} \biggl\{
-\frac{1}{2}\biggl[
c_{\delta}^{2} \frac{M_{H^{0}}^{2}}{M_{W}^{2}} 
\ln \biggl(\frac{M_{H^0}^{2}}{M_{W}^{2}}\biggr)
\\
&&\qquad
+4s_{\delta}^{2} \frac{M_{K^{0}}^{2}}{M_{W}^{2}} 
\ln \biggl(\frac{M_{K^0}^{2}}{M_{W}^{2}}\biggr)
+s_{\delta}^{2} \frac{M_{H^{\pm}}^{2}M_{Z}^{2}}{M_{W}^{4}} 
\ln \biggl(\frac{M_{H^\pm}^{2}}{M_{Z}^{2}}\biggr)
\biggr]
\nonumber\\
&&
-s_{\delta}^{2}\frac{M_{H^{0}}^{2} M_{H^{\pm}}^{2}}{2M_{W}^{4}} 
\ln \biggl(\frac{M_{H^\pm}^{2}}{M_{H^{0}}^{2}}\biggr)
- c_{\delta}^{2} \frac{2M_{K^{0}}^{2}M_{H^{\pm}}^{2}}{M_{W}^{4}} 
\ln \biggl(\frac{M_{H^\pm}^{2}}{M_{K^{0}}^{2}}\biggr)
\biggr\} 
 \; ,
\nonumber
\end{eqnarray}
for $M_{H^{0}} \ll M_{K^{0}} \ll M_{H^{\pm}}$. 
Thus the scalar contribution to 
$\Delta r_{triplet}$ in this case is very large, 
and it grows with the scalar masses. On the other hand, when 
the mass splitting between either pair of the three scalar 
masses is small (case (a) and (b) and their generalization), 
the scalar contributions grow with the mass 
splitting\cite{Chen:2003fm,Toussaint:1978zm,Senjanovic:1978ee}, 
\begin{eqnarray}\label{cc2}
\Delta r_{triplet}^{S} &\rightarrow &
\frac{\alpha}{4\pi \hat{s}_{\theta}^{2}} 
 \biggl\{
-\frac{1}{2}\biggl[
c_{\delta}^{2} \frac{M_{H^{0}}^{2}}{M_{W}^{2}} 
\ln \biggl(\frac{M_{H^0}^{2}}{M_{W}^{2}}\biggr)
\\
&&\qquad
+4s_{\delta}^{2} \frac{M_{K^{0}}^{2}}{M_{W}^{2}} 
\ln \biggl(\frac{M_{K^0}^{2}}{M_{W}^{2}}\biggr)
+s_{\delta}^{2} \frac{M_{H^{\pm}}^{2}M_{Z}^{2}}{M_{W}^{4}} 
\ln \biggl(\frac{M_{H^\pm}^{2}}{M_{Z}^{2}}\biggr)
\biggr]
\nonumber\\
&& \qquad +\frac{5}{72}\biggl[
s_{\delta}^{2}\frac{
\bigl(M_{H^{\pm}}^{2}-M_{H^{0}}^{2}\bigr)}{M_{H^{0}}^{2}} 
+ 4c_{\delta}^{2} \frac{ 
\bigl(M_{H^{\pm}}^{2} - M_{K^{0}}^{2}\bigr)}{M_{K^{0}}^{2}}
\biggr]
\biggr\} 
\nonumber \; ,
\end{eqnarray}
for $M_{H^{0}} \simeq M_{K^{0}} \simeq M_{H^{\pm}}$, and, 
\begin{eqnarray}\label{cc3}
\Delta r_{triplet}^{S} & \rightarrow &
\frac{\alpha}{4\pi \hat{s}_{\theta}^{2}} \biggl\{
-\frac{1}{2}\biggl[
c_{\delta}^{2} \frac{M_{H^{0}}^{2}}{M_{W}^{2}} 
\ln \biggl(\frac{M_{H^0}^{2}}{M_{W}^{2}}\biggr)
\\
&&\qquad
+4s_{\delta}^{2} \frac{M_{K^{0}}^{2}}{M_{W}^{2}} 
\ln \biggl(\frac{M_{K^0}^{2}}{M_{W}^{2}}\biggr)
+s_{\delta}^{2} \frac{M_{H^{\pm}}^{2}M_{Z}^{2}}{M_{W}^{4}}  
\ln \biggl(\frac{M_{H^\pm}^{2}}{M_{Z}^{2}}\biggr)
\biggr]
\nonumber\\
&&
-s_{\delta}^{2}\frac{M_{H^{0}}^{2}M_{H^{\pm}}^{2}}{2M_{W}^{4}} 
\ln \biggl(\frac{M_{H^\pm}^{2}}{M_{H^{0}}^{2}}\biggr)
+ \frac{5}{18}c_{\delta}^{2} 
\frac{\bigl(M_{H^{\pm}}^{2} - M_{K^{0}}^{2}\bigr)}{M_{K^{0}}^{2}}
\biggr\}
\nonumber \; ,
\end{eqnarray}
for $M_{H^{0}} \ll M_{K^{0}} \simeq M_{H^{\pm}}$. 
Cancellations can occur in this case among contributions 
from different scalar fields, leading to the viability 
of a heavier neutral Higgs boson than is allowed in the SM\cite{Forshaw:2003kh}.

The non-decoupling property of the triplet is seen in Eq.~(\ref{cc1}), 
(\ref{cc2}) and (\ref{cc3}). Because $\Delta r_{triplet}$ 
depends quadratically on the 
scalar masses\cite{Lynn:1990zk,Passarino:1990nu,Gunion:1990dt,Pomarol:1993mu,Forshaw:2001xq}, 
the scalars must be included in any effective field theory 
analysis of low energy physics.

The scalar potential of the model with a $SU(2)_{L}$ triplet 
and an $SU(2)_{L}$ doublet is given by 
the following\cite{Forshaw:2003kh}:
\begin{equation}
V(H,\Phi) = \mu_{1}^{2} \bigl|H\bigr|^{2} 
+ \frac{1}{2} \mu_{2}^{2} \Phi^{\dagger} \Phi
+ \lambda_{1} \bigl|H\bigr|^{4} 
+ \frac{1}{4} \lambda_{2} \bigl|\Phi^{\dagger}\Phi\bigr|^{2}  
+ \frac{1}{2} \lambda_{3} \bigl|H\bigr|^{2} \Phi^{\dagger}\Phi 
+ \lambda_{4} \Phi^{\alpha}_{U} H^{\dagger} \sigma^{\alpha} H
\; ,
\end{equation}
where $\sigma^{\alpha}$ denotes the Pauli matrices, and 
\begin{equation}
\Phi_{U} = U^{\dagger} \Phi, \quad 
U = \frac{1}{\sqrt{2}} \left(
\begin{array}{ccc}
1 & -i & 0\\
0 & 0 & \sqrt{2}\\
-1 & -i & 0
\end{array}\right)
\; . 
\end{equation}
From the minimization conditions (see \ref{minimize}),
\begin{equation}
\frac{\partial V}{\partial \eta^{0}} 
\bigg|_{<H>,<\Phi>} = 
\frac{\partial V}{\partial \phi^{0}}
\bigg|_{<H>,<\Phi>} =0,
\end{equation} 
we obtain the following conditions,
\begin{eqnarray}
4\mu_{2}^{2} t_{\delta} + \lambda_{2}v^{2}t_{\delta}^{3}
+2\lambda_{3}v^{2}t_{\delta}-4\lambda_{4}v & = & 0
\label{min1}\\
\mu_{1}^{2} + \lambda_{1}v^{2} + \frac{1}{8} 
\lambda_{3}v^{2}t_{\delta}^{2} 
-\frac{1}{2}\lambda_{4}vt_{\delta} & = & 0
\; .\label{min2}
\end{eqnarray}
The two mixing angles, $\gamma$ and $\delta$, in the 
neutral and charged Higgs sectors defined 
in Eqs.~(\ref{neutralmix}) and (\ref{chargedmix}), are solutions 
to the following two equations\cite{Forshaw:2003kh},
\begin{eqnarray}
0 & = & 
\lambda_{4} v + \tan\delta \biggl[
\mu_{1}^{2} - \mu_{2}^{2} + \lambda_{1} v^{2} 
- \frac{1}{2}\lambda_{3} v^{2} 
+ \lambda_{4} v^{\prime} 
- \lambda_{2} v^{\prime}
+ \frac{1}{2} \lambda_{3} v^{\prime}
\\&& \qquad 
- \lambda_{4} v \tan\delta
\biggr]
\nonumber
\label{cond1}\\
0 &=&
-\lambda_{4}v + \lambda_{3}vv^{\prime}+\tan\gamma 
\biggl[
\mu_{1}^{2} - \mu_{2}^{2} + 3 \lambda_{1} v^{2}
- \frac{1}{2} \lambda_{3} v^{2} 
-\lambda_{4}v^{\prime}
-3\lambda_{2} v^{\prime 2}
\\
&&\qquad
+ \frac{1}{2} \lambda_{3} v^{\prime 2} 
+ \lambda_{4} v \tan\gamma
- \lambda_{3} vv^{\prime}\tan\gamma
\biggr] \; ,\label{cond2} \nonumber
\end{eqnarray}
which are obtained by minimizing the scalar potential. 
In terms of the parameters in the scalar potential, 
the masses of the four scalar fields are given 
by\cite{Forshaw:2003kh},
\begin{eqnarray}
M_{H^{\pm}}^{2} & = & 
\mu_{2}^{2} + \lambda_{2} v^{2} \tan^{2}\delta 
+ \lambda_{4} v \tan\delta
+ \frac{1}{2} \lambda_{3} v^{2}
\\
M_{H^{0}}^{2} & = & 
\mu_{1}^{2} + 3 \lambda_{1} v^{2}
+ \lambda_{3} v^{2} \tan\delta
\biggl(\frac{1}{2} \tan\delta-\tan\gamma \biggr) 
+\lambda_{4} v \biggl( \tan\gamma-\tan\delta \biggr)
\\
M_{K^{0}}^{2} & = & 
\mu_{2}^{2} + 3 \lambda_{2} v^{2} \tan^{2}\delta
-\lambda_{4}v \tan\gamma
+\frac{1}{2}\lambda_{3} v^{2}\biggl( 1+2\tan\delta\tan\gamma \biggr) \;.   
\end{eqnarray}
This model has six parameters in the scalar sector, 
$\bigl(\mu_{1}^{2},\mu_{2}^{2},
\lambda_{1},\lambda_{2},\lambda_{3},\lambda_{4}\bigr)$. Equivalently, 
we can choose $\bigl(M_{H^{0}},M_{K^{0}},M_{H^{\pm}},
v,\tan\delta,\tan\gamma \bigr)$ as the independent parameters.
Two of these six parameters, $v$ and $\tan\delta$, contribute to the 
gauge boson masses. 

When turning off the couplings between the doublet and the 
triplet in the scalar potential, $\lambda_{3}=\lambda_{4}=0$, 
the triplet could still acquire a VEV, $v^{\prime} \sim 
\sqrt{\frac{-2\mu_{2}^{2}}{\lambda_{2}}}$, provided that $\mu_{2}^{2}$ 
is negative. Since we have not observed any light scalar experimentally 
up to the EW scale, the triplet mass which is roughly of order $\mu_{2}$ 
has to be at least of the EW scale, $v \lesssim \mu_{2}$. 
This is problematic because the VEV of a real triplet 
only contributes to $M_{W}$ but not to $M_Z$, 
which then results in a contribution of order $\mathcal{O}(1)$ 
to the $\rho$ parameter, due to  
the relation, $\rho=1+4\frac{v^{\prime2}}{v^2}$. 
For $\mu_{2}$ greater than $v$, the EW symmetry is broken at a high scale. 
In order to avoid these problems, the parameter 
$\mu_{2}^{2}$ thus has to be positive so that the triplet does 
not acquire a VEV via this mass term when $\lambda_{4}$ is turned off. 
Once the coupling $\lambda_{3}$ is turned on while keeping $\lambda_{4}=0$, 
the term $\lambda_{3} |H|^{2} \Phi^{\dagger}\Phi$ effectively 
plays the role of the mass term for $\Phi$ and for $H$. Thus, similar 
to the reasoning given above, for $\mu_{2} 
\sim v$, the coupling $\lambda_{3}$ has to be positive so that 
it does not induce a large triplet VEV. 
For simplicity, consider the case when there is no mixing in 
the neutral Higgs sector, $\gamma=0$. In this case, 
when the mixing in the charged sector approaches zero, 
$\delta \rightarrow 0$, the masses $M_{K^{0}}$ and 
$M_{H^{\pm}}$ approach infinity, and their difference 
$M_{K^{0}}^{2}-M_{H^{\pm}}^{2}$ approaches zero. 
The contribution due to the new scalars thus vanishes, 
and only the lightest neutral Higgs contributes to 
$\Delta r_{triplet}$. Even though the contribution 
due to the new scalars vanishes,    
$\Delta r_{triplet}$ does not approach $\Delta r_{SM}$. 
This is because in the TM case, four input parameters are needed, 
while in the SM case three inputs are needed. There is no continuous 
limit that takes one case to the other\cite{Passarino:1990xx,Lynn:1990zk,Passarino:1990nu,Gunion:1990dt,Pomarol:1993mu,Forshaw:2001xq}.

One way to achieve the $\delta \rightarrow 0$ limit 
is to take the mass parameter $\mu_{2}^{2} 
\rightarrow \infty$ while keeping the parameter 
$\lambda_{4}$ finite. Eq.~(\ref{cond1}) then dictates that 
$\mu_{2}^{2} \tan\delta \sim \lambda_{4} v \sim v^{2}$.  
However, satisfying Eq.~(\ref{cond2}) requires that  
$\lambda_{4} v \sim \lambda_{3}vv^{\prime} = \lambda_{3}v^{2} \tan\beta$. 
As $\lambda_{4}v \sim v^{2}$, this condition implies that 
the dimensionless coupling constant $\lambda_{3}$ has to scale 
as $\mu_{2}^{2}/v^{2}$, which approaches infinity as 
$\delta \rightarrow 0$. 
This can also be seen from Eq.~(\ref{neutralmix1}). As there is 
no mixing in the neutral sector, 
\begin{equation}
\frac{\partial^{2} V}
{\partial \phi^{0} \partial \eta^{0}}
= \frac{1}{2} \lambda_{3} v^{2} t_{\delta} - \lambda_{4} v
 = 0 \; ,
\end{equation} 
the condition  
\begin{equation}
\tan \delta = \frac{2\lambda_{4}}{\lambda_{3}v} \; 
\end{equation}
then follows. 
So, in the absence of the neutral mixing, $\gamma=0$, 
in order to take the charged mixing angle $\delta$ to zero 
while holding $\lambda_{4}$ fixed, one has to take 
$\lambda_{3}$ to infinity. In other words, for the 
triplet to decouple requires a dimensionless 
coupling constant $\lambda_{3}$ to become strong, 
leading to the breakdown of the perturbation theory.

Alternatively, the neutral mixing angle $\gamma$ can approach zero 
by taking $\mu_{2}^{2} \rightarrow \infty$ while keeping 
$\lambda_{3}$ and $\lambda_{4}$ fixed. In this case, 
the minimization condition, 
\begin{equation}
4\mu_{2}^{2}t_{\delta} + \lambda_{2} v^{2} t_{\delta}^{3} 
+ 2 \lambda_{3} v^{2} t_{\delta} - 4 \lambda_{4} v = 0 \;, 
\end{equation}
where $t_{\delta} \equiv \tan\delta$, 
implies that the charged mixing angle  
$\delta$ has to approach zero. This again corresponds to the 
case where the custodial symmetry is restored, by which 
we mean that the triplet VEV vanishes, $v^{\prime} = 0$. 
In this case, severe fine-tuning is needed in order 
to satisfied the condition given in 
Eq.~(\ref{cond2}).    
Another way to get $\delta \rightarrow 0$ is to have $\lambda_{4} 
\rightarrow 0$, which trivially satisfies Eq.~(\ref{cond1}). 
This can also be seen from Eq.~\ref{chargemix2}, 
\begin{equation}
\frac{\partial^{2}V}
{\partial \eta^{+}\partial \phi^{-}}=\lambda_{4}v = 0 \; .
\end{equation} 
Eq.~(\ref{cond2}) then gives $\lambda_{4}\cot\delta 
\sim \lambda_{3} v$. So for small $\lambda_{3}$, the masses of these 
additional scalar fields are of the weak scale, $M_{K^{0}} \sim 
M_{H^{\pm}} \sim v$. This corresponds to a case when the custodial symmetry 
is restored. So unless one imposes by hand such symmetry to forbid 
$\lambda_{4}$, four input parameters are always needed in the renormalization.
If there is a symmetry which makes $\lambda_{4} = 0$ (to all orders), 
say, $\Phi \rightarrow -\Phi$, then there are only three input parameters 
needed. So the existence of such a symmetry is crucial when one-loop radiative 
corrections are concerned.

\begin{figure}[bh!]
\begin{center}
\includegraphics[scale=0.6]{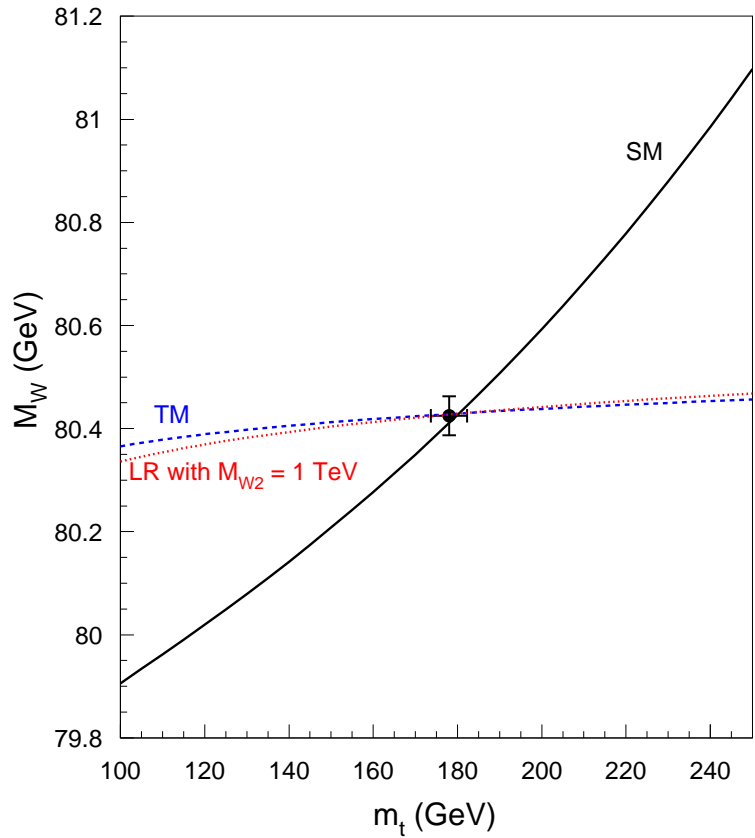}
\vspace*{-0.2cm}
\caption[]{Prediction for the $W$ mass as a function of the
top quark mass in the SM, TM and the LR model. 
The data point represents the experimental values with 
$1\sigma$ error bars\cite{ewpewwg}. For the SM, we include the 
complete contributions from top and bottom quarks, the SM Higgs boson 
with $M_{H^{0}}=120$ GeV, and the gauge bosons. 
For the TM and the LR model, we include only the top 
quark contribution and the absolute normalization 
is fixed so that the curves intersect the data point. The 
$W_{2}$ boson mass is chosen to be $M_{W_{2}}=1$ TeV 
in the LR model.}
\label{fg:mwmt}
\end{center}
\end{figure}
\begin{figure}[b!]
\begin{center}
\includegraphics[scale=0.4]{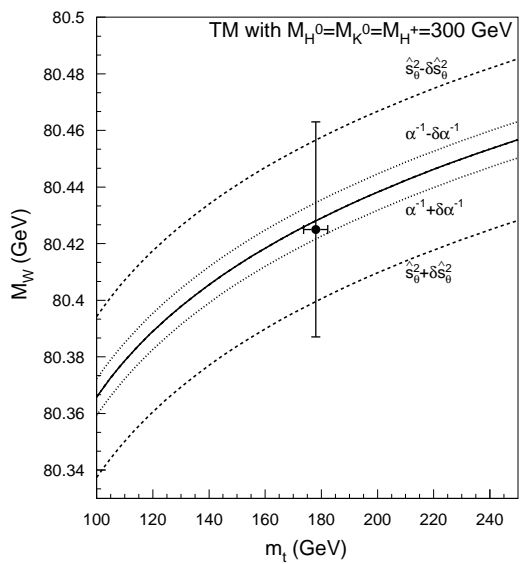}
\vspace*{-0.2cm}
\caption[]{Prediction for the $W$ mass in the TM as a function of the
top quark mass for scalar masses $M_{H^{0}}=M_{K^{0}}=M_{H^{\pm}}
=300$ GeV, 
with $\alpha(M_Z)$ and $\hat{s}_{\theta}$ varying within 
their $1\sigma$ limits\cite{pdg04,ewpewwg} 
$\alpha(M_{Z})^{-1} = 128.91\pm 0.0392$ 
and $\hat{s}_{\theta}^{2}=0.2315\pm 0.000314$. The solid 
curve indicates the prediction with $\alpha(M_Z)$ and $\hat{s}_{\theta}$ 
taking the experimental central values, $\alpha(M_{Z})^{-1} = 
128.91$ and $\hat{s}_{\theta}^{2} = 0.2315$.  
The area bounded by the short dashed (dotted) curves indicates the prediction 
with $\alpha(M_{Z})^{-1} = 128.91$ ($\hat{s}_{\theta}^{2}=0.2315$)
and $\hat{s}_{\theta}^{2}$ ($\alpha(M_{Z})$) varying 
with its $1\sigma$ limits.  
The data point represents 
the experimental values with $1\sigma$ error bars\cite{ewpewwg}.}
\label{fg:mwmterr}
\end{center}
\end{figure}
\begin{figure}[b!]
\begin{center}
\includegraphics[scale=0.3]{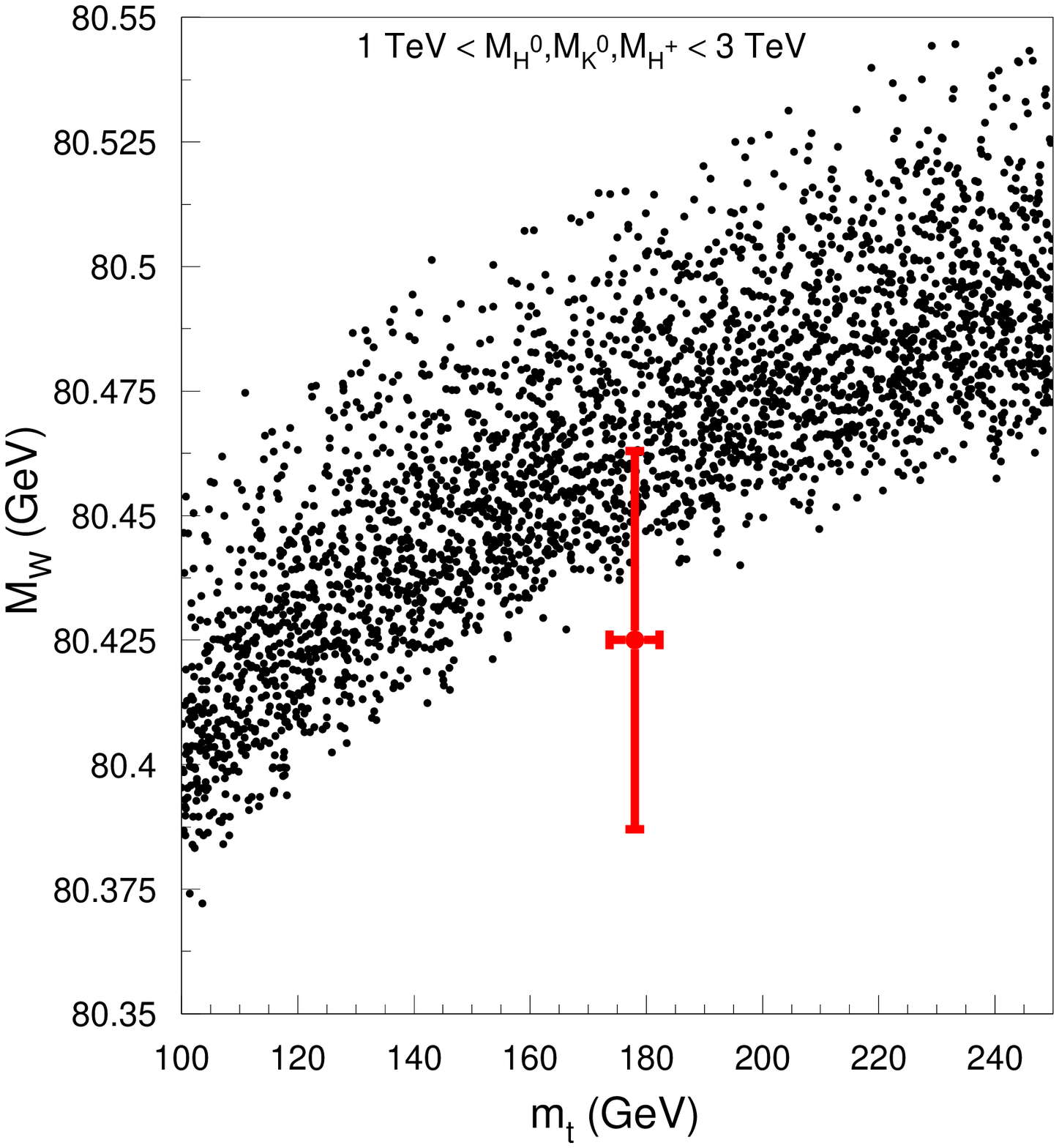}
\includegraphics[scale=0.3]{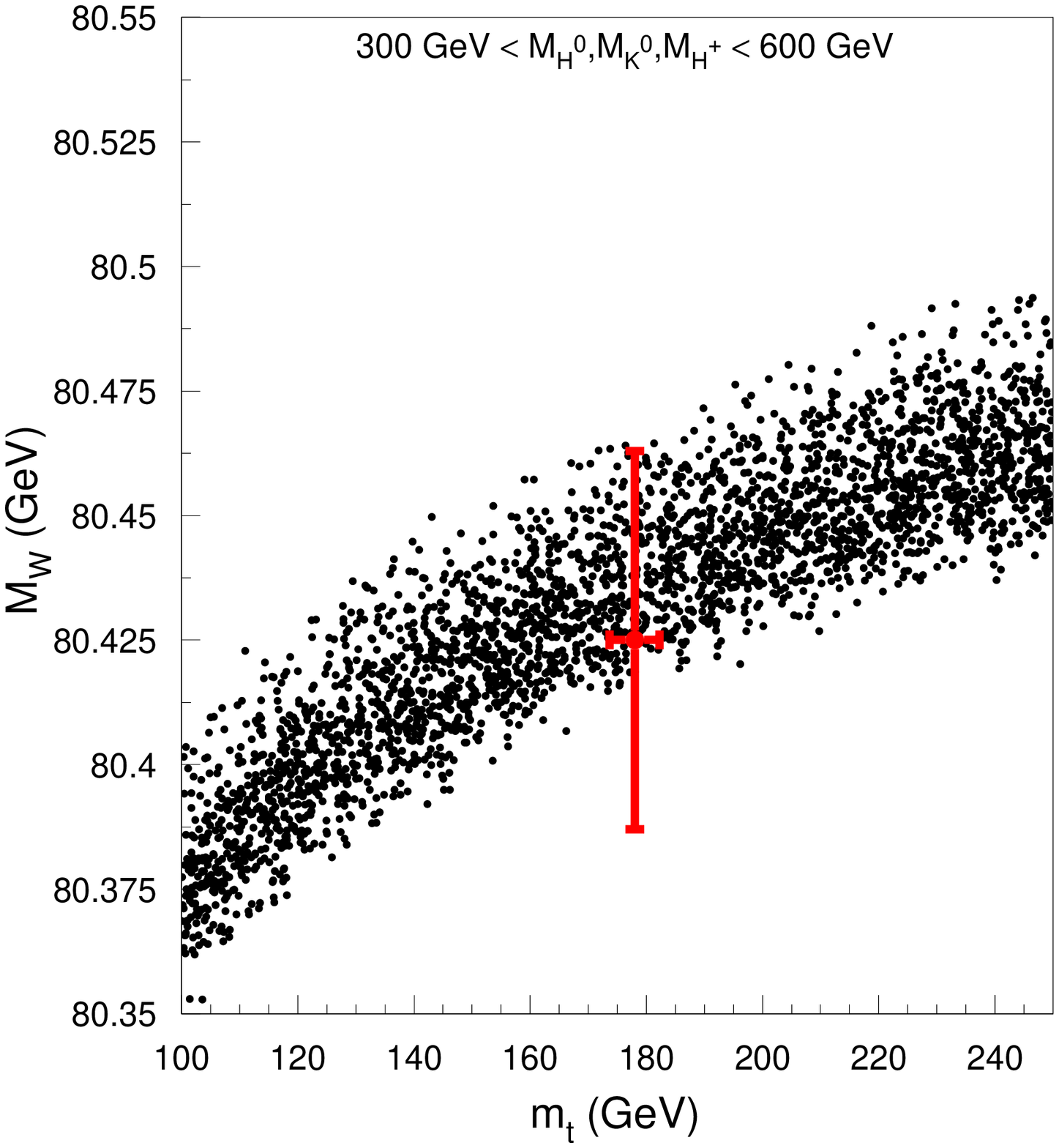}
\includegraphics[scale=0.3]{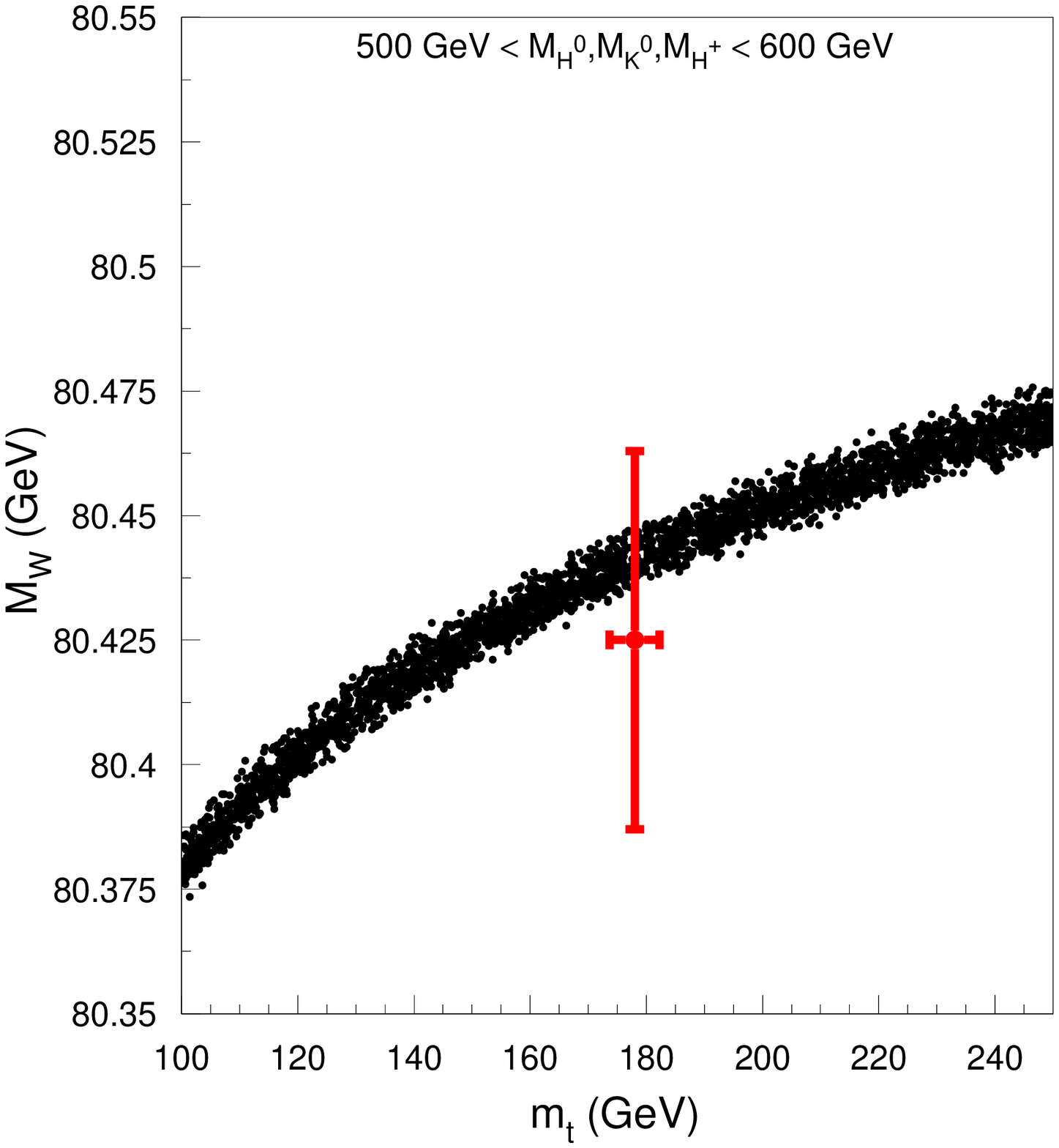}
\caption[]{Prediction for the $W$ mass in the TM as a function of the
top quark mass for scalar masses, $M_{H^{0}}$, $M_{K^{0}}$ and 
$M_{H^{\pm}}$, varying independently between 
(a) $1-3$ TeV, (b) $300-600$ GeV, and (c) $500-600$ GeV. 
The data point represents 
the experimental values with $1\sigma$ error bars\cite{ewpewwg}.}
\label{fg:mwmt1}
\end{center}
\end{figure}
\begin{figure}[b!]
\begin{center}
\includegraphics[scale=0.3]{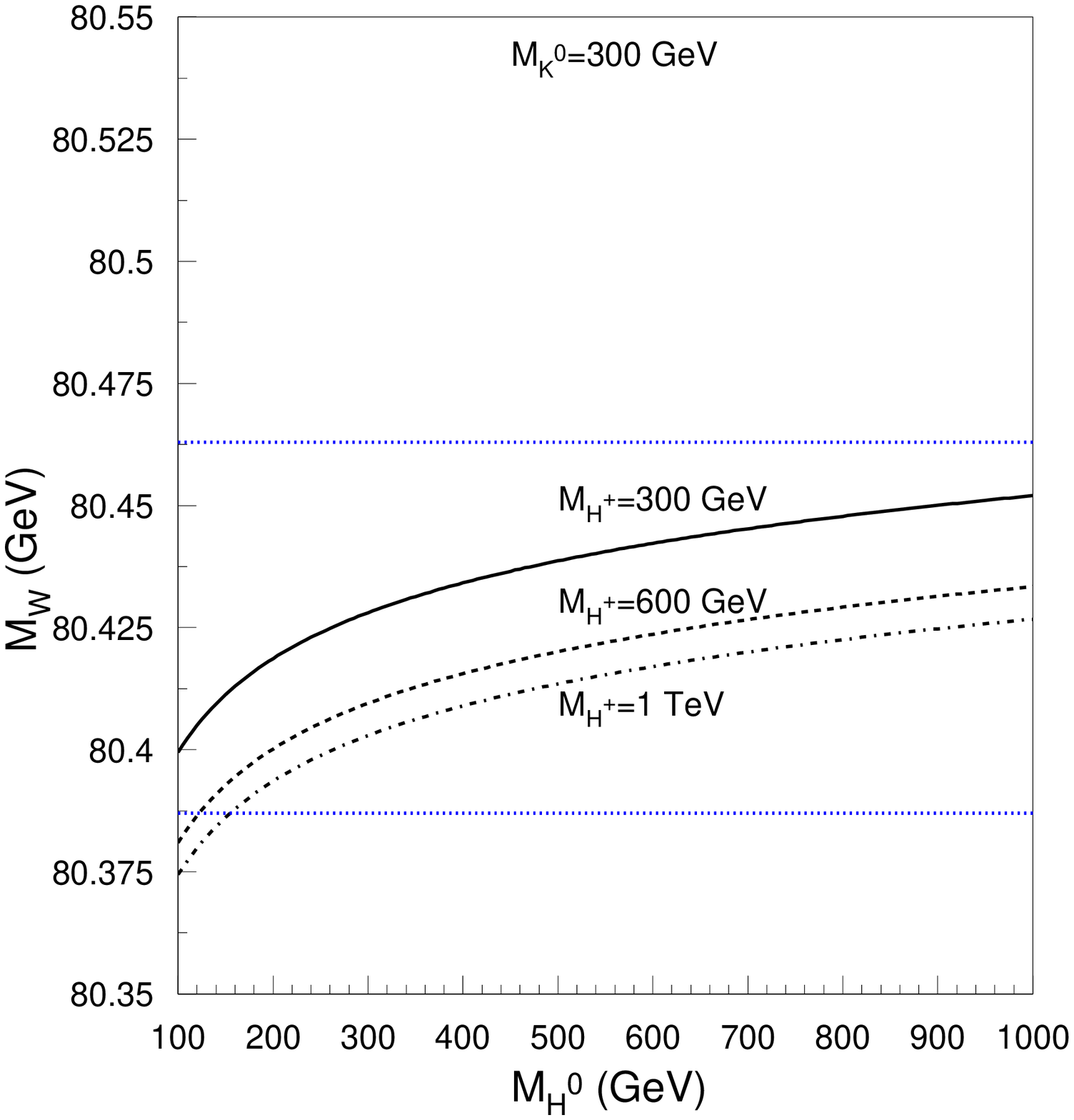}
\includegraphics[scale=0.3]{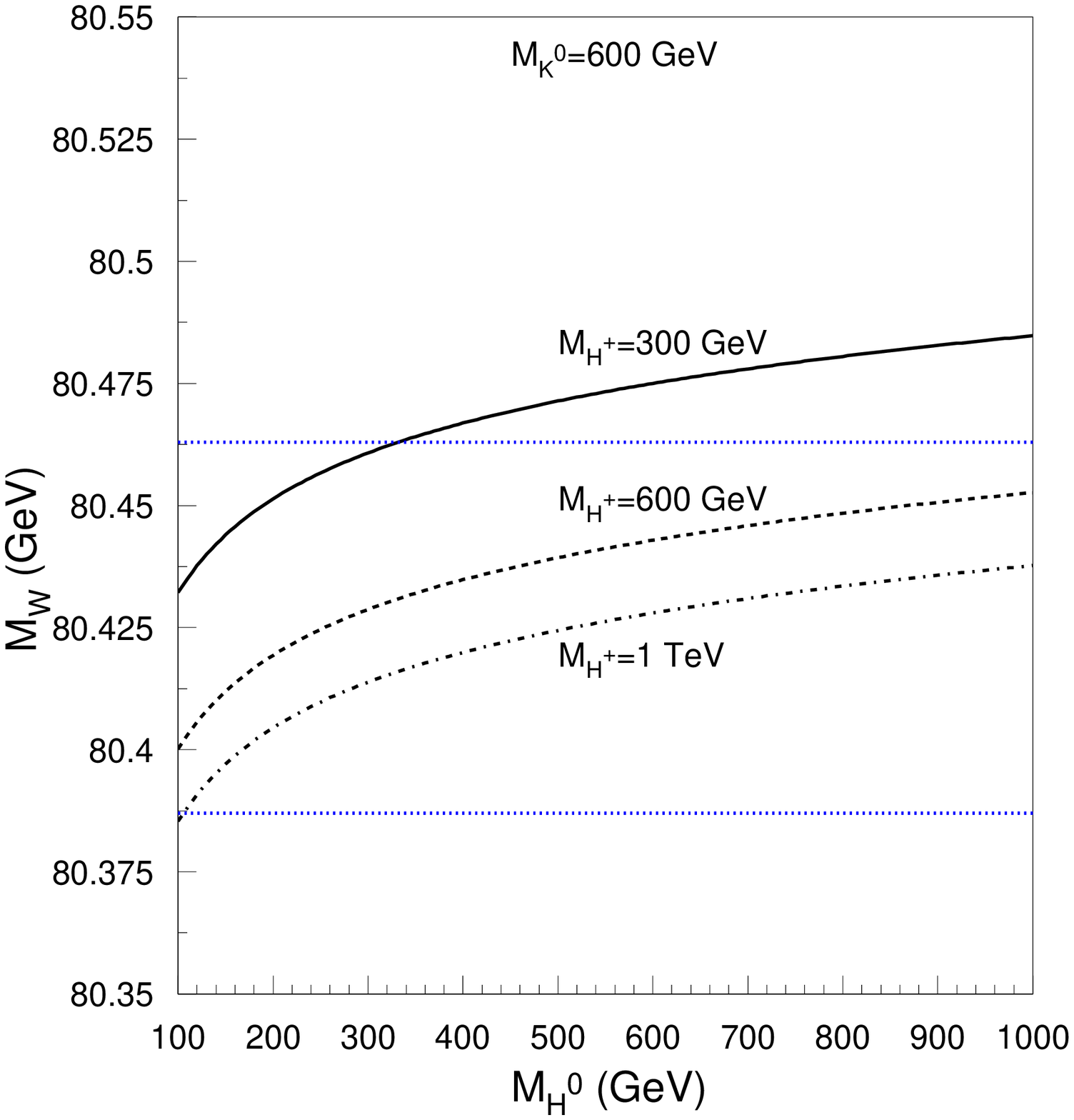}
\includegraphics[scale=0.3]{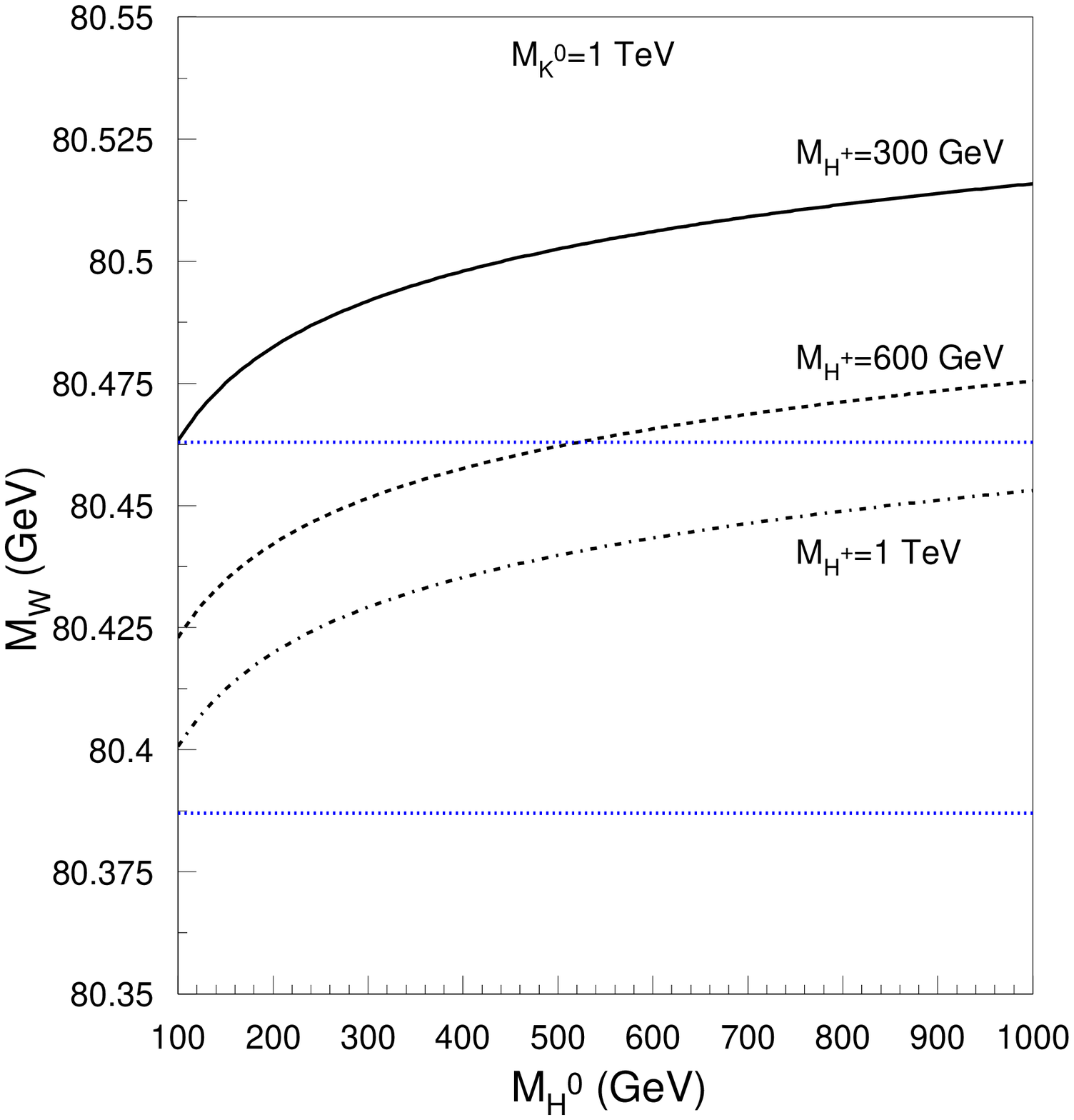}
\caption[]{Prediction for the $W$ mass in the TM as a function of the
lightest neutral Higgs boson mass, $M_{H^{0}}$, 
for various values of $M_{K^{0}}$ and $M_{H^{\pm}}$. The area  
bounded by the two horizontal lines is the $1\sigma$ 
allowed region for $M_{W}$\cite{ewpewwg}.}
\label{fg:mwmh}
\end{center}
\end{figure}

\section{Results}\label{results}

The previous section has presented analytic results for the 
triplet model, demonstrating that the dependence of the $W$ mass 
on the top quark mass is logarithmic, while the dependence on the 
scalar masses is quadratic. A dramatic change in the behavior 
of the $W$ mass is also observed in the $SU(2)_{L} \times 
SU(2)_{R}$ model\cite{Czakon:1999ue,Czakon:1999ha,Czakon:2002wm}. 
For comparison with the triplet model, we summarize the results 
of the left-right model in Appendix~\ref{lrmodel}. In this case, the 
dependence of the $W$ mass on the top quark mass is weakened from that of the 
SM since it depends on $m_{t}^{2}/M_{W_{2}}^{2}$, where
$M_{W_{2}}$ is the heavy charged gauge boson mass of the left-right model. 

The dependence of the $W$ mass on the top quark mass, $m_{t}$, 
in the case of the SM, the model with a triplet Higgs, 
and the minimal left-right model, are shown in Fig.~\ref{fg:mwmt}. 
For the SM, we include the complete contributions from 
top and bottom quarks, the SM Higgs boson with 
$M_{H^{0}}=120$ GeV, and the gauge bosons. In this case, 
the $m_{t}$ dependence in the prediction for $M_W$ is quadratic. 
 The range of values for the input parameter 
$m_{t}$ that give a prediction for $M_{W}$ consistent with the experimental  
$1\sigma$ limits\cite{ewpewwg}, $M_{W}=80.425 \pm 0.0666$ GeV, 
is very narrow. It coincides with 
the current experimental limits\cite{ewpewwg}, $178 \pm 4.3$ GeV.
For the triplet model and the LR model, 
we include only the top quark contribution. 
As we have shown in Sec.~\ref{higgstrip}, the prediction 
for $M_W$ in the triplet model depends on $m_{t}$ only 
logarithmically. 
In our numerical result for the left-right model, 
we have used $\bigl( G_{\mu}, \, \alpha(M_{Z}), \, 
\hat{s}_{\theta}, \, M_{Z}, \, M_{W_{2}} \bigr)$ in the gauge 
sector, in addition to $m_{t}$ in the fermion sector, to predict $M_{W}$. 
Here we have identified $W_{1}$ and $Z_{1}$ as the $W$ and $Z$ 
bosons in the SM and 
consequently $M_{W} = M_{W_{1}}$ and $M_{Z}=M_{Z_{1}}$.
In this case, the $m_{t}$ dependence in the 
prediction for $M_{W}$ is similarly softer because the top quark 
contributions are suppressed by a heavy scale, $M_{W_{2}}$.
In the triplet model and the left-right model, the range 
of $m_{t}$ that gives a prediction for $M_{W}$ 
consistent with the experimental value is thus much larger,  
ranging from $m_{t}=120$ to $250$ GeV. The presence of the triplet 
Higgs thus dramatically changes the $m_{t}$ dependence in $M_{W}$.  
This is clearly demonstrated in Fig.~\ref{fg:mwmt} 
by the almost flat curves of the triplet and left-right 
symmetric models, contrary to that of the SM, which is very 
sensitive to $m_{t}$.   
In Fig.~\ref{fg:mwmterr}, 
we show the prediction for $M_{W}$ as a function of $m_{t}$ in the triplet 
model, with  $\alpha(M_Z)$ and $\hat{s}_{\theta}$ varying within 
the $1\sigma$ limits\cite{pdg04,ewpewwg}, 
$\alpha(M_{Z})^{-1} = 128.91\pm 0.0392$ 
and $\hat{s}_{\theta}^{2}=0.2315\pm 0.000314$. 
We find that the prediction for $M_{W}$ is very sensitive to 
the input parameters $\alpha(M_{Z})$ and $\hat{s}_{\theta}$.

The complete contributions from the top and bottom quarks and 
the SM gauge bosons, as well as all four scalar fields in the triplet model  
are included in Fig.~\ref{fg:mwmt1} 
and \ref{fg:mwmh}. 
We have also included the box and vertex corrections. 
In Fig.~\ref{fg:mwmt1}, 
we show the prediction in the triplet model 
for $M_{W}$ as a function of $m_{t}$, 
allowing $M_{H^{0}}$, $M_{H^{\pm}}$ and $M_{K^{0}}$ to vary 
independently between $1-3$ TeV, $300-600$ GeV and $500-600$ GeV. 
Interestingly, for all scalar masses in the range of $1-3$ TeV, 
the prediction for $M_{W}$ in the TM model still 
agrees with the experimental $1\sigma$ limits. 
Fig.~\ref{fg:mwmh} shows the prediction for $M_{W}$ 
as a function of $M_{H^{0}}$ for various values of $M_{K^{0}}$ 
and $M_{H^{\pm}}$. For small $M_{K^{0}}^{2}-M_{H^{\pm}}^{2}$, the lightest  
neutral Higgs boson mass can range from $M_{H^{0}}=100$ GeV to a TeV 
and still satisfy the experimental prediction for $M_{W}$.  
This agrees with our conclusion in Sec.~\ref{higgstrip} that   
to minimize the scalar contribution to $\Delta r_{triplet}$, 
the mass splitting $M_{K^{0}}^{2}-M_{H^{\pm}}^{2}$ 
has to be small and that when the mass splitting is small, 
cancellations can occur between the contributions of the lightest neutral  
Higgs and those of the additional scalar fields. This has new 
important implications for the Higgs searches.

\section{Conclusion}
\label{conc}

We have considered the top quark contribution to muon decay at one loop in 
the SM and in two models with $\rho\ne 1$ at tree level:  the SM
with an addition real scalar triplet and the minimal left-right 
model. In these new models, because the $\rho$ parameter is no longer
 equal to one at the tree level, a fourth input parameter is 
required in a consistent renormalization scheme.  
These models illustrate a general feature that the $m_{t}$ 
dependence in the radiative corrections $\Delta r_{triplet}$ becomes 
logarithmic, contrary to the case of the SM where $\Delta r_{SM}$ 
depends on $m_t$ quadratically. One therefore loses the 
prediction for $m_{t}$ from radiative corrections. On the 
other hand, due to  cancellations between the contributions to 
the radiative corrections from the SM Higgs and the triplet, 
a Higgs mass $M_{H^{0}}$ as large as a few TeV is allowed 
by the $W$ mass measurement. 
We emphasize that by taking the triplet mass to infinity, 
one does not recover the SM. This is due to 
the fact that the triplet scalar field is non-decoupling, and 
it implies that the one-loop electroweak results 
cannot be split into a SM contribution plus a piece which 
vanishes as the scale of new physics becomes much larger than the weak 
scale. This fact has been overlooked by most analyses in the littlest 
Higgs model\cite{Csaki:2002qg,Hewett:2002px}, 
and correctly including the effects of the triplet can dramatically 
change the conclusion on the viability of the  
model\cite{Chen:2003fm,Chen:2004ig}. 
Such non-decoupling effect has been pointed out in the two Higgs doublet 
model\cite{Toussaint:1978zm}, 
left-right symmetric model\cite{Senjanovic:1978ee}, 
and the littlest Higgs model\cite{Chen:2003fm}. 
It has {\it not} been discussed before in the model with a triplet. 
We comment that the non-decoupling observed in these examples do not 
contradict with the common knowledge that in GUT models heavy scalars 
decouple. These two cases are fundamentally different because in GUT models, 
heavy scalar fields do not acquire VEV that break the EW symmetry, 
while in cases where non-decoupling is observed, heavy scalar fields 
{\it do} acquire VEV that breaks the EW symmetry.  The quadratic dependence 
in scalar masses in the triplet model can be easily understood physically. In SM  
with only the Higgs doublet present, the quadratic scalar mass contribution is 
protected by the tree level custodial symmetry, and thus the Higgs mass contribution is 
logarithmic at one-loop. This is the well-known screening theorem by 
Veltman\cite{Veltman:1977kh}.  As the custodial symmetry is broken in the SM at one-loop 
due to the mass splitting between the top and bottom quarks, the two-loop Higgs 
contribution is quadratic. In models with a triplet Higgs, as the custodial symmetry 
is broken already at the tree level, there is no screening theorem 
that protects the quadratic scalar mass dependence from appearing.
Our results demonstrate the importance of performing the 
renormalization correctly according to the EW structure of the new models.

\section{Acknowledgments}

This manuscript has been authored by Brookhaven Science Associates, 
LLC under Contract No. DE-AC02-76CH1-886 with the U.S. 
Department of Energy. The United States Government retains, 
and the publisher, by accepting the article for publication, 
acknowledges, a world-wide license to publish or reproduce 
the published form of this manuscript, or allow others to do so, 
for the United States Government purpose. 
T.K. thanks W. Marciano for useful discussions.

\appendix
\section{Contributions of the top loop}\label{loop}

We summarize below the leading contributions due to the 
SM top loop to the self-energies of the gauge 
bosons\cite{Chen:2003fm}, where the definitions of 
the Passarino-Veltman functions utilized below are given 
in\cite{Chen:2003fm}. 
\begin{eqnarray}
\Pi^{WW}(M_{W}^{2}) &=& 
-\frac{3 \alpha}{4\pi s_{\theta}^{2}} \biggl[
A_{0}(m_{t}^{2}) + 
m_{b}^{2} B_{0}(M_{W}^{2},m_{b}^{2},m_{t}^{2})\\
&& -
M_{W}^{2} B_{1}(M_{W}^{2},m_{b}^{2},m_{t}^{2})
- 2 B_{22}(M_{W}^{2},m_{b}^{2},m_{t}^{2}) \biggr]
\nonumber\\
\Pi^{WW} (0) & = & -\frac{3 \alpha}{16\pi s_{\theta}^{2}}  \, \cdot \,
m_{t}^{2} \biggl[ 1 + 2 \ln \biggl( \frac{Q^{2}}{m_{t}^{2}} \biggr) \biggr] 
\\
\Pi^{ZZ} (M_{Z}^{2}) & = & -\frac{3 \alpha}{8\pi s_{\theta}^{2} 
c_{\theta}^{2}} 
\biggl[ 
\biggl(\biggl( \frac{1}{2} - \frac{4}{3}s_{\theta}^{2} \biggr)^{2} 
+ \frac{1}{4} \biggr)  
h_{1}(m_{t}^{2})\\
&&
-\frac{8}{3} s_{\theta}^{2} \biggl(
 1-\frac{4}{3}s_{\theta}^{2} \biggr) h_{2}(m_{t}^{2})
\biggr]\nonumber
\\
\Pi_{\gamma\gamma}^{\prime}(0) & = &
\frac{4\alpha}{9\pi} 
\ln \biggl(\frac{Q^{2}}{m_{t}^{2}}\biggr) \\
\Pi^{\gamma Z} (M_{Z}^{2}) &=& -\frac{\alpha}{\pi s_{\theta}c_{\theta}}
\biggl( \frac{1}{2} - \frac{4}{3} s_{\theta}^{2} \biggr) M_{Z}^{2}
\biggl[ \frac{1}{3}  \ln \biggl(\frac{Q^{2}}{m_{t}^{2}}\biggr)
-2 I_{3}\biggl(\frac{M_{Z}^{2}}{m_{t}^{2}}\biggr)\biggr]
\end{eqnarray}
where
\begin{eqnarray}
h_{1}(m_{t}^{2}) & = &  2m_{t}^{2}
 \biggl[ \ln \biggl( \frac{Q^{2}}{m_{t}^{2}} \biggr) 
- \frac{1}{3} I_{1} \biggl( \frac{M_{Z}^{2}}{m_{t}^{2}} \biggr) \biggr]
\\
h_{2}(m_{t}^{2}) & = & m_{t}^{2} 
\biggl[ I_{1}\biggl(\frac{M_{Z}^{2}}{m_{t}^{2}}\biggr) 
- \ln \biggl(\frac{Q^{2}}{m_{t}^{2}}\biggr)
\biggr]\quad .
\end{eqnarray} 
The integrals are defined as, 
\begin{eqnarray}
I_1(a)&=&\int_0^1 dx 
\ln\biggl(1-ax(1-x)\biggr)\nonumber \\
I_3(a)&=&\int_0^1 dx x (1-x) 
\ln \biggl(1-ax(1-x)\biggr)\nonumber \; .
\end{eqnarray}
Here $s_{\theta}$ is defined in the on-shell scheme (Eq.~(\ref{swos})) 
for the SM and as the effective weak mixing angle 
(Eq.~(\ref{swdef})) for the TM and LR model.

\section{Contributions of the scalars in a model with a triplet Higgs}
\label{scalar}

The complete contributions to various two-point functions that appear in 
$\Delta r_{triplet}$ are given below, where the scalar and fermion 
contributions are given in Ref.~\refcite{Blank:1997qa,Chen:2003fm}, 
and we have taken the SM gauge boson contributions from 
Ref.~\refcite{Hollik:1993cg}.
\begin{eqnarray}
\Pi_{WW}(0)-\Pi_{WW}(M_{W}) & = & 
-\frac{3 \alpha}{16\pi \hat{s}_{\theta}^{2}}  \, \cdot \,
m_{t}^{2} \biggl[ 1 + 2 \ln \biggl( \frac{Q^{2}}{m_{t}^{2}} \biggr) \biggr]
\\
& & +\frac{3 \alpha}{4\pi \hat{s}_{\theta}^{2}} \biggl[
A_{0}(m_{t}^{2}) + 
m_{b}^{2} B_{0}(M_{W}^{2},m_{b}^{2},m_{t}^{2}) \nonumber\\
& & \qquad 
-M_{W}^{2} B_{1}(M_{W}^{2},m_{b}^{2},m_{t}^{2})
- 2 B_{22}(M_{W}^{2},m_{b}^{2},m_{t}^{2}) \biggr]
\nonumber\\
& & + {\alpha\over 4\pi \hat{s}_{\theta}^{2}}\biggl\{
s_\delta^2H(M_{H^0}, M_{H^\pm})+c_\delta^2H(M_{H^0}, M_W)\nonumber \\
&& \qquad +
4c_\delta^2H(M_{K^0}, M_{H^\pm})+4s_\delta^2H(M_{K^0}, M_W)\nonumber \\
&& \qquad +
s_\delta^2H(M_Z, M_{H^\pm})+c_\delta^2H(M_Z, M_W)\biggr\}
\nonumber\\
& & + {\alpha\over 4 \pi \hat{s}_{\theta}^{2}}M_W^2
\biggl[ {s_\delta^2 c_\delta^2\over \hat{c}_{\theta}^{2}} 
\biggl(B_0(0,M_Z,M_{H^\pm})- B_0(M_{W},M_Z,M_{H^\pm}) \biggr)
\nonumber\\
& & \qquad 
+{(s_\delta^2-\hat{s}_\theta^2)^2\over \hat{c}_{\theta}^{2}}
\biggl( B_0(0,M_Z,M_W) - B_0(M_{W},M_Z,M_W) \biggr)
\nonumber \\ 
&& \qquad 
+ \hat{s}_\theta^2 
\biggl(B_0(0,0,M_W) - B_0(M_W,0,M_W) \biggr)
\nonumber\\
& & \qquad 
+c_\delta^2 
\biggl( B_0(0,M_{H^0}, M_W) - B_0(M_W,M_{H^0}, M_W) \biggr)
\nonumber\\
& & \qquad 
+4 s_\delta^2 
\biggl(B_0(0,M_{K^0},M_W) - B_0(M_W,M_{K^0},M_W) \biggr) 
\biggr]
\nonumber \\
& & 
+ \frac{\alpha}{4\pi \hat{s}_{\theta}^{2}} \biggl[
\hat{c}_{\theta}^{2} \biggl( A_{1}(0,M_{Z},M_{W}) - A_{1}(M_W,M_Z,M_W) \biggr)
\nonumber\\
& & \qquad 
+ \hat{s}_{\theta}^{2} \biggl(
A_{1}(0,0,M_W) - A_{1}(M_W,0,M_W) \biggr)
\nonumber\\
& & \qquad 
- 2\hat{c}_{\theta}^{2} H(M_Z,M_W) 
- 2\hat{s}_{\theta}^{2} H(0,M_W)
\biggr]\nonumber \; ,
\\
&&\nonumber
\\&&\nonumber
\\&&\nonumber
\\&&\nonumber
\\&&\nonumber
\\&&\nonumber
\\&&\nonumber
\\&&\nonumber
\\&&\nonumber
\\&&\nonumber
\end{eqnarray}

\begin{eqnarray}
\Pi_{\gamma Z} (M_{Z}) & = & 
-\frac{\alpha}{\pi \hat{s}_{\theta}\hat{c}_{\theta}}
\biggl( \frac{1}{2} - \frac{4}{3} \hat{s}_{\theta}^{2} \biggr) M_{Z}^{2}
\biggl[ \frac{1}{3}  \ln \biggl(\frac{Q^{2}}{m_{t}^{2}}\biggr)
-2 I_{3}\biggl(\frac{M_{Z}^{2}}{m_{t}^{2}}\biggr)\biggr]
\\
& & + 
\frac{\alpha}{4\pi \hat{s}_{\theta}\hat{c}_{\theta}}
\biggl[
2(c_{\delta}^{2}-\hat{s}_{\theta}^{2} 
+ \hat{c}_{\theta}^{2})B_{22}(M_Z,M_{H^\pm},M_{H^\pm})
\nonumber\\
&&\qquad
+ 2 (s_{\delta}^2-\hat{s}_{\theta}^2+\hat{c}_{\theta}^2) B_{22}(M_Z,M_W,M_W) 
\nonumber\\
& & \qquad 
+ (\hat{s}_{\theta}^2-\hat{c}_{\theta}^2-c_{\delta}^2)A(M_{H^\pm}) 
+ (\hat{s}_{\theta}^2-\hat{c}_{\theta}^2-s_{\delta}^2)A(M_{W}) \biggr]
\nonumber\\
& & + 
\frac{\alpha}{4\pi}\bigl(2M_{W}^{2})
\frac{\hat{s}_{\theta}^2-s_{\delta}^2}
{\hat{s}_{\theta}\hat{c}_{\theta}} B_{0}(M_Z,M_W,M_W)
\nonumber\\
& & 
-\frac{\alpha}{4\pi \hat{s}_{\theta}^{2}} 
\biggl[ \hat{s}_{\theta}\hat{c}_{\theta} A_{1}(M_Z,M_W,M_W)
+ 2 \hat{c}_{\theta}\hat{s}_{\theta} A_{2}(M_W)
\nonumber\\
&&\qquad
+ 2 \hat{s}_{\theta}\hat{c}_{\theta} B_{22}(M_Z,M_W,M_W)
\biggr] \; ,
\nonumber
\end{eqnarray}

\begin{eqnarray}
\Pi_{\gamma \gamma}^{\prime}(0) & = & 
\frac{\alpha}{\pi} \biggl[
\frac{4}{9} 
\ln \biggl(\frac{Q^{2}}{m_{t}^{2}}\biggr) +
 \frac{1}{12} \ln \biggl(
\frac{Q^2}{M_{H^\pm}^{2}} \biggr) - \frac{3}{4} 
\ln \biggl( \frac{Q^{2}}{M_{W}^{2}}\biggr)
- \frac{1}{6} \biggr] \; ,
\end{eqnarray}

\begin{eqnarray}
\Pi_{\gamma Z}(0) & = & \frac{\alpha}{4\pi}
\biggl[
\frac{(\hat{s}_{\theta}^{2}-s_{\delta}^{2})}{\hat{s}_{\theta}\hat{c}_{\theta}} 
2M_{W}^{2} B_{0}(0,M_{W},M_{W})
- \frac{\hat{c}_{\theta}}{\hat{s}_{\theta}}A_{1}(0,M_W,M_W)
\\
&&\qquad
- 2 \frac{\hat{c}_{\theta}}{\hat{s}_{\theta}} A_{2}(M_W) 
- 2 \frac{\hat{c}_{\theta}}{\hat{s}_{\theta}} B_{22}(0,M_W,M_W)
\biggr] \; ,
\nonumber
\end{eqnarray}
where
\begin{eqnarray}
H(m_{1},m_{2}) &=& -B_{22}(0,m_{1},m_{2}) + B_{22}(M_W,m_{2},m_{2})
\; ,\\
A_{1}(p,m_{1},m_{2}) & = &
-A_{0}(m_{1})-A_{0}(m_{2})-(m_{1}^{2}+m_{2}^{2}+4p^{2})
B_{0}(p,m_{1},m_{2})
\\&&\qquad
-10B_{22}(p,m_{1},m_{2})
+2(m_{1}^{2}+m_{2}^{2}-\frac{p^{2}}{3}) \; ,
\nonumber\\
A_{2}(m)&=& 3A_{0}(m)-2m^{2} \; ,
\end{eqnarray}
and $\overline{s}_{\theta}$ is defined in Eq.~(\ref{swdef}). 

To extract the dependence on the masses of the lightest neutral Higgs, 
$M_{H^{0}}$, and the extra scalar fields, 
$M_{K^0}$ and $M_{H^\pm}$, we first 
note that,  in the limit $\delta m^{2} \ll m_{1}^{2}$, 
\begin{eqnarray}
B_{0}(p,m_{1},m_{2}) & = & 
\ln \biggl(\frac{Q^{2}}{m_{1}^{2}}\biggr)
+ \frac{1}{6} \frac{p^{2}}{m_{1}^{2}} 
- \frac{1}{2} \frac{\delta m^{2}}{m_{1}^{2}}
+ \mathcal{O}\biggl( (\delta m^{2})^{2}, \; 
\biggl(\frac{p^{2}}{m_{1}^{2}}\biggr)^{2} \biggr)
\\
B_{22}(p,m_{1},m_{2}) & = & 
\frac{1}{2} m_{1}^{2} \biggl[ 1 + 
\ln \biggl(\frac{Q^{2}}{m_{1}^{2}}\biggr) \biggr]
-\frac{1}{12} p^{2} \ln \biggl( \frac{Q^{2}}{m_{1}^{2}}
\biggr) -\frac{1}{72}\frac{p^{4}}{m_{1}^{2}}
\\
&&
+ \biggl[ \frac{1}{4} \ln \biggl(\frac{Q^{2}}{m_{1}^{2}} \biggr) 
+ \frac{5}{72}\frac{p^{2}}{m_{1}^{2}}\biggr] \delta m^{2} 
+ \mathcal{O}\biggl( (\delta m^{2})^{2}, \; 
\biggl(\frac{p^{2}}{m_{1}^{2}}\biggr)^{2} \biggr)\nonumber \\
&& 
+ ( \mbox{terms with no scalar dependence})\nonumber
\nonumber \; ,
\end{eqnarray}
where we have defined $\delta m^{2} = m_{2}^{2} - m_{1}^{2}$ 
and assumed that $p^{2} \ll m_{1}^{2}$. Using 
these relations, we then have,
\begin{eqnarray}
H(m_{1},m_{2}) & = & 
\frac{5}{72} \frac{M_{W}^{2}}{m_{1}^{2}} \delta m^{2}
- \frac{1}{72}  \frac{M_{W}^{4}}{m_{1}^{2}} 
+ \mathcal{O}\biggl( (\delta m^{2})^{2}, \; 
\biggl(\frac{M_{W}^{2}}{m_{1}^{2}}\biggr)^{2} \biggr) 
\label{h:eq}\\
&& 
+ ( \mbox{terms with no scalar dependence}) \; ,
\nonumber
\end{eqnarray}
\begin{eqnarray}
B_{0}(0,m_{1},m_{2}) - B_{0}(M_{W},m_{1},m_{2}) & = &
-\frac{1}{6}\frac{M_{W}^{2}}{m_{1}^{2}}
+ \mathcal{O}\biggl( (\delta m^{2})^{2}, \; 
\biggl(\frac{M_{W}^{2}}{m_{1}^{2}}\biggr)^{2} \biggr) 
\label{b:eq}
\end{eqnarray}
On the other hand, in the limit $m_{1} \gg m_{2}$, we get,
\begin{eqnarray}
B_{0}(p,m_{1},m_{2}) &=& \biggl(
1 + \ln \biggl(\frac{Q^{2}}{m_{1}^{2}}\biggr) 
\biggr)\biggl(1+\frac{m_{2}^{2}}{m_{1}^{2}}\biggr)
+\frac{1}{2}\frac{p^{2}}{m_{1}^{2}} 
\\
&&
+ \mathcal{O}\biggl( \biggl( \frac{m_{2}^{2}}{m_{1}^{2}} 
\biggr)^{2}, \; 
\biggl(\frac{p^{2}}{m_{1}^{2}}\biggr)^{2} \biggr)\nonumber
\\
B_{22}(p,m_{1},m_{2})&=&
m_{1}^{2} \biggl( 
\frac{3}{8} + \frac{2}{3} \frac{m_{2}^{2}}{m_{1}^{2}} - \frac{1}{18}
\frac{p^{2}}{m_{1}^{2}} - \frac{m_{2}^{2}}{12p^{2}} \biggr)
\\
&&
- \biggl( \frac{1}{4}
+\frac{1}{6}\frac{m_{2}^{2}}{m_{1}^{2}}
+\frac{m_{2}^{2}}{12p^{2}}-\frac{p^{2}}{12m_{1}^{2}}\biggr)
m_{1}^{2} \ln m_{1}^{2}
\nonumber\\
&&
- \biggl(
\frac{m_{2}^{2}}{12m_{1}^{2}}-\frac{m_{2}^{2}}{12p^{2}}
\biggr)m_{1}^{2}\ln m_{2}^{2}
 +  \biggl( \frac{1}{4}m_{1}^{2}+\frac{1}{4} m_{2}^{2}
 -\frac{p^{2}}{12}\biggr)\ln Q^{2}
\nonumber\\
&&
+ \mathcal{O} \biggl( \biggl(\frac{m_{2}^{2}}{m_{1}^{2}}\biggr), 
\;  \biggl(\frac{p^{2}}{m_{1}^{2}}\biggr) \biggr)
+ \; ( \mbox{terms with no scalar dependence})
\nonumber\; ,
\end{eqnarray}
which gives,
\begin{eqnarray}
H(m_{1},m_{2})&=&
-\frac{m_{1}^{2}m_{2}^{2}}{12M_{W}^{2}} \biggl[ 1+ \ln \biggl(
\frac{m_{1}^{2}}{m_{2}^{2}}\biggr) \biggr]
+\mathcal{O} \biggl( \biggl(\frac{m_{2}^{2}}{m_{1}^{2}}\biggr), 
\;  \biggl(\frac{p^{2}}{m_{1}^{2}}\biggr) \biggr) 
\\
&&+
 \; ( \mbox{terms with no scalar dependence})
\nonumber
\end{eqnarray}
\begin{eqnarray}
B_{0}(0,m_{1},m_{2})-B_{0}(M_{W},m_{1},m_{2})&=&
-\frac{1}{2}\frac{M_{W}^{2}}{m_{1}^{2}}
+  
\mathcal{O}\biggl( \biggl( \frac{m_{2}^{2}}{m_{1}^{2}} 
\biggr)^{2}, \; 
\biggl(\frac{M_{W}^{2}}{m_{1}^{2}}\biggr)^{2} \biggr)
\label{b0:eq2}
\end{eqnarray}

The two-point function $\Pi_{\gamma Z}(0)$ does not 
have any scalar dependence, and the function $\Pi_{\gamma 
\gamma}^{\prime}(0)$ depends on the scalar mass 
only logarithmically, 
\begin{equation}
\Pi_{\gamma\gamma}^{\prime}(0) \rightarrow
\frac{\alpha}{12\pi}\ln\biggl(\frac{Q^{2}}{M_{H^{\pm}}^{2}}\biggr)
\; .
\end{equation}
The scalar dependence in the function 
$\Pi_{\gamma Z}(M_{Z})$ is,
\begin{eqnarray}
\Pi_{\gamma Z}(M_{Z}) & \rightarrow & 
\frac{\alpha}{4\pi \hat{s}_{\theta} \hat{c}_{\theta}} (\hat{s}_{\theta}^{2}
-\hat{c}_{\theta}^{2}-c_{\delta}^{2}) 
\biggl[ A_{0}(M_{H^{\pm}})
-2B_{22}(M_{Z},M_{H^{\pm}},M_{H^{\pm}})
\biggr]
\\
& = & 
\frac{\alpha}{4\pi \hat{s}_{\theta} \hat{c}_{\theta}} 
(\hat{s}_{\theta}^{2}
-\hat{c}_{\theta}^{2}-c_{\delta}^{2}) M_{Z}^{2}
\biggl[
 \frac{1}{12} \ln \biggl(
\frac{M_{H^{\pm}}^{2}}{Q^{2}}\biggr) 
- \frac{1}{72} \frac{M_{Z}^{2}}{M_{H^{\pm}}^{2}}\biggr] \; , 
\nonumber
\end{eqnarray}
thus the dependence is also logarithmic. On the other hand, in the 
function $\Pi_{WW}(0)-\Pi_{WW}(M_{W})$, the scalar dependence 
is given by,
\begin{eqnarray}
\Pi_{WW}(0)-\Pi_{WW}(M_{W})
&\rightarrow & 
{\alpha\over 4\pi \hat{s}_\theta^2}\biggl\{
s_\delta^2H(M_{H^0}, M_{H^\pm})+c_\delta^2H(M_{H^0}, M_W)
\label{square}\\
&&  
+ 4c_\delta^2H(M_{K^0}, M_{H^\pm})+4s_\delta^2H(M_{K^0}, M_W)
+s_\delta^2H(M_Z, M_{H^\pm})\biggr\}
\nonumber\\
& & + {\alpha\over 4 \pi \hat{s}_\theta^2}M_W^2
\biggl[ {s_\delta^2 c_\delta^2\over \hat{c}_\theta^2} 
\biggl(B_0(0,M_Z,M_{H^\pm})- B_0(M_{W},M_Z,M_{H^\pm}) \biggr)
\nonumber\\
& & \qquad 
+c_\delta^2 
\biggl( B_0(0,M_{H^0}, M_W) - B_0(M_W,M_{H^0}, M_W) \biggr)
\nonumber\\
& & \qquad 
+4 s_\delta^2 
\biggl(B_0(0,M_{K^0},M_W) - B_0(M_W,M_{K^0},M_W) \biggr) 
\biggr] \; .
\nonumber
\end{eqnarray}
From Eqs.~(\ref{b:eq}) and (\ref{b0:eq2}), we know that the contributions 
from the terms in the square brackets of Eq.~(\ref{square}) are logarithmic. 
Thus the only possible quadratic dependence comes from terms in the curly 
brackets. We consider the following three limits, assuming that all 
scalar masses are much larger than $M_{W}$ and $M_{Z}$:

\begin{itemize}

\item[(a)] $M_{H^{0}} \simeq M_{K^{0}} \simeq M_{H^{\pm}}$: 
In this case, the leading order scalar dependence is given by,
\begin{eqnarray}
\Pi_{WW}(0)-\Pi_{WW}(M_{W}) & \rightarrow &
\frac{\alpha}{4\pi \hat{s}_{\theta}^{2}} 
\biggl\{ -\frac{1}{2}\biggl[
c_{\delta}^{2} M_{H^{0}}^{2} 
\ln \biggl(\frac{M_{H^0}^{2}}{M_{W}^{2}}\biggr)
\\
&&\qquad
+4s_{\delta}^{2} M_{K^{0}}^{2} 
\ln \biggl(\frac{M_{K^0}^{2}}{M_{W}^{2}}\biggr)
+s_{\delta}^{2} \frac{M_{H^{\pm}}^{2}M_{Z}^{2}}{M_{W}^{2}} 
\ln \biggl(\frac{M_{H^{\pm}}^{2}}{M_{Z}^{2}}\biggr)
\biggr]
\nonumber\\
&&+\frac{5}{72}\biggl[
s_{\delta}^{2}\frac{M_{W}^{2}}{M_{H^{0}}^{2}} 
\bigl(M_{H^{\pm}}^{2}-M_{H^{0}}^{2}\bigr)
+ 4c_{\delta}^{2} \frac{M_{W}^{2}}{M_{K^{0}}^{2}} 
\bigl(M_{H^{\pm}}^{2} - M_{K^{0}}^{2}\bigr)
\biggr]
\biggr\}
\nonumber \; .
\end{eqnarray}
So the dominant scalar contribution to 
$\Delta r_{triplet}^{S}$ in this case is given by,
\begin{eqnarray}
\Delta r_{triplet}^{S} &\rightarrow &
\frac{\alpha}{4\pi \hat{s}_{\theta}^{2}} 
 \biggl\{
-\frac{1}{2}\biggl[
c_{\delta}^{2} \frac{M_{H^{0}}^{2}}{M_{W}^{2}} 
\ln \biggl(\frac{M_{H^0}^{2}}{M_{W}^{2}}\biggr)
\\
&&\qquad
+4s_{\delta}^{2} \frac{M_{K^{0}}^{2}}{M_{W}^{2}} 
\ln \biggl(\frac{M_{K^0}^{2}}{M_{W}^{2}}\biggr)
+s_{\delta}^{2} \frac{M_{H^{\pm}}^{2}M_{Z}^{2}}{M_{W}^{4}} 
\ln \biggl(\frac{M_{H^\pm}^{2}}{M_{Z}^{2}}\biggr)
\biggr]
\nonumber\\
&& \qquad +\frac{5}{72}\biggl[
s_{\delta}^{2}\frac{
\bigl(M_{H^{\pm}}^{2}-M_{H^{0}}^{2}\bigr)}{M_{H^{0}}^{2}} 
+ 4c_{\delta}^{2} \frac{ 
\bigl(M_{H^{\pm}}^{2} - M_{K^{0}}^{2}\bigr)}{M_{K^{0}}^{2}}
\biggr]
\biggr\} 
\nonumber \; .
\end{eqnarray}

\item[(b)] $M_{H^{0}} \ll M_{K^{0}} \simeq M_{H^{\pm}} $: In this limit, 
the leading scalar dependence becomes,
\begin{eqnarray}
\Pi_{WW}(0)-\Pi_{WW}(M_{W}) & \rightarrow &
\frac{\alpha}{4\pi \hat{s}_{\theta}^{2}} \biggl\{
-\frac{1}{2}\biggl[
c_{\delta}^{2} M_{H^{0}}^{2} 
\ln \biggl(\frac{M_{H^0}^{2}}{M_{W}^{2}}\biggr)
\\
&&\qquad
+4s_{\delta}^{2} M_{K^{0}}^{2} 
\ln \biggl(\frac{M_{K^0}^{2}}{M_{W}^{2}}\biggr)
+s_{\delta}^{2} \frac{M_{H^{\pm}}^{2}M_{Z}^{2}}{M_{W}^{2}} 
\ln \biggl(\frac{M_{H^\pm}^{2}}{M_{Z}^{2}}\biggr)
\biggr]
\nonumber\\
&&
-s_{\delta}^{2}\frac{M_{H^{0}}^{2}}{2M_{W}^{2}} 
\ln \biggl(\frac{M_{H^\pm}^{2}}{M_{H^{0}}^{2}}\biggr)
M_{H^{\pm}}^{2}
+ \frac{5}{18}c_{\delta}^{2} \frac{M_{W}^{2}}{M_{K^{0}}^{2}} 
\bigl(M_{H^{\pm}}^{2} - M_{K^{0}}^{2}\bigr)
\biggr\}
\nonumber \; .
\end{eqnarray}
The leading scalar contribution to 
$\Delta r_{triplet}^{S}$ is,
\begin{eqnarray}
\Delta r_{triplet}^{S} & \rightarrow &
\frac{\alpha}{4\pi \hat{s}_{\theta}^{2}} \biggl\{
-\frac{1}{2}\biggl[
c_{\delta}^{2} \frac{M_{H^{0}}^{2}}{M_{W}^{2}} 
\ln \biggl(\frac{M_{H^0}^{2}}{M_{W}^{2}}\biggr)
\\
&&\qquad
+4s_{\delta}^{2} \frac{M_{K^{0}}^{2}}{M_{W}^{2}} 
\ln \biggl(\frac{M_{K^0}^{2}}{M_{W}^{2}}\biggr)
+s_{\delta}^{2} \frac{M_{H^{\pm}}^{2}M_{Z}^{2}}{M_{W}^{4}}  
\ln \biggl(\frac{M_{H^\pm}^{2}}{M_{Z}^{2}}\biggr)
\biggr]
\nonumber\\
&&
-s_{\delta}^{2}\frac{M_{H^{0}}^{2}M_{H^{\pm}}^{2}}{2M_{W}^{4}} 
\ln \biggl(\frac{M_{H^\pm}^{2}}{M_{H^{0}}^{2}}\biggr)
+ \frac{5}{18}c_{\delta}^{2} 
\frac{\bigl(M_{H^{\pm}}^{2} - M_{K^{0}}^{2}\bigr)}{M_{K^{0}}^{2}}
\biggr\}
\nonumber \; .
\end{eqnarray}

\item[(c)] $M_{H^{0}} \ll M_{K^{0}} \ll M_{H^{\pm}} $: In this limit, 
the leading scalar dependence becomes,
\begin{eqnarray}
\Pi_{WW}(0)-\Pi_{WW}(M_{W}) & \rightarrow &
\frac{\alpha}{4\pi \hat{s}_{\theta}^{2}} \biggl\{
-\frac{1}{2}\biggl[
c_{\delta}^{2} M_{H^{0}}^{2} 
\ln \biggl(\frac{M_{H^0}^{2}}{M_{W}^{2}}\biggr)
\\
&&\qquad
+4s_{\delta}^{2} M_{K^{0}}^{2} 
\ln \biggl(\frac{M_{K^0}^{2}}{M_{W}^{2}}\biggr)
+s_{\delta}^{2} \frac{M_{H^{\pm}}^{2}M_{Z}^{2}}{M_{W}^{2}} 
\ln \biggl(\frac{M_{H^\pm}^{2}}{M_{Z}^{2}}\biggr)
\biggr]
\nonumber\\
&&
-s_{\delta}^{2}\frac{M_{H^{0}}^{2}}{2M_{W}^{2}} 
\ln \biggl(\frac{M_{H^\pm}^{2}}{M_{H^{0}}^{2}}\biggr)
M_{H^{\pm}}^{2}
- c_{\delta}^{2} \frac{2M_{K^{0}}^{2}}{M_{W}^{2}} 
\ln \biggl(\frac{M_{H^\pm}^{2}}{M_{K^{0}}^{2}}\biggr)
M_{H^{\pm}}^{2}
\biggr\} \; ,
\nonumber
\end{eqnarray}
The leading scalar contribution to 
$\Delta r_{triplet}^{S}$ is thus given by,
\begin{eqnarray}
\Delta r_{triplet}^{S} &\rightarrow&
\frac{\alpha}{4\pi \hat{s}_{\theta}^{2}} \biggl\{
-\frac{1}{2}\biggl[
c_{\delta}^{2} \frac{M_{H^{0}}^{2}}{M_{W}^{2}} 
\ln \biggl(\frac{M_{H^0}^{2}}{M_{W}^{2}}\biggr)
\\
&&\qquad
+4s_{\delta}^{2} \frac{M_{K^{0}}^{2}}{M_{W}^{2}} 
\ln \biggl(\frac{M_{K^0}^{2}}{M_{W}^{2}}\biggr)
+s_{\delta}^{2} \frac{M_{H^{\pm}}^{2}M_{Z}^{2}}{M_{W}^{4}} 
\ln \biggl(\frac{M_{H^\pm}^{2}}{M_{Z}^{2}}\biggr)
\biggr]
\nonumber\\
&&
-s_{\delta}^{2}\frac{M_{H^{0}}^{2} M_{H^{\pm}}^{2}}{2M_{W}^{4}} 
\ln \biggl(\frac{M_{H^\pm}^{2}}{M_{H^{0}}^{2}}\biggr)
- c_{\delta}^{2} \frac{2M_{K^{0}}^{2}M_{H^{\pm}}^{2}}{M_{W}^{4}} 
\ln \biggl(\frac{M_{H^\pm}^{2}}{M_{K^{0}}^{2}}\biggr)
\biggr\} 
 \; .
\nonumber
\end{eqnarray}
For the case $M_{H^{0}} \ll M_{H^{\pm}} \ll M_{K^{0}}$, make the 
replacement, $ \ln \biggl( \frac{M_{H^\pm}^{2}}{M_{K^{0}}^{2}}\biggr) 
\longleftrightarrow \ln\biggl(\frac{M_{K^0}^{2}}{M_{H^{\pm}}^{2}}\biggr)$.

\end{itemize}

\section{The Left-Right Symmetric Model}
\label{lrmodel}

As our second example to show that new physics does not decouple from the 
SM at one-loop, we consider the left-right (LR) symmetric 
model\cite{Pati:1974yy,Mohapatra:1974hk,Mohapatra:1974gc,Senjanovic:1975rk} 
which is defined by the gauge group,
\begin{equation}
SU(2)_L\times SU(2)_R\times U(1)_{B-L}\quad .
\end{equation}
The minimal left-right symmetric model contains a scalar bi-doublet, 
$\Phi$, and two $SU(2)$ triplets, $\Delta_L$ and $\Delta_R$. 
We assume that the scalar potential is arranged such 
that the Higgs fields obtain the following VEV's:
\begin{eqnarray}
\Phi \sim & (1/2,1/2,0) & \sim 
\left( \begin{array}{cc}
\kappa & 
\\
 & \kappa^\prime
\end{array}\right)
\\
\Delta_{L} \sim & (1,0,2) & \sim
\left( \begin{array}{cc}
0 & 0
\\
v_{L} & 0
\end{array}\right)
\\
\Delta_{R} \sim & (0,1,2) & \sim
\left( \begin{array}{cc}
0 & 0 
\\
v_{R} & 0
\end{array}\right) \; ,
\end{eqnarray}
where the quantum numbers of these Higgs fields under $SU(2)_{L}$, 
$SU(2)_{R}$ and $U(1)_{B-L}$  
are given inside the parentheses.  
The VEV $v_{R}$ breaks the $SU(2)_R \times U(1)_{B-L}$ symmetry  
down to $U(1)_{Y}$ of the SM, while the bi-doublet VEV's 
$\kappa$ and $\kappa^\prime$ break 
the electroweak symmetry; the VEV $v_{L}$ may be relevant for generating 
neutrino masses\cite{Chen:2004ww,chen:2003a,chen:2004a,chen:2000a,chen:2001a,chen:2002a,chen:2004b,chen:2004c,chen:2005a}.  After the symmetry breaking,
there are two charged gauge bosons, $W_1$ and $W_2$, two heavy neutral
gauge bosons, $Z_1$ and $Z_2$, and the massless photon.  We will assume
that $W_1$ and $Z_1$ are the lighter gauge bosons and obtain roughly their 
SM values after the symmetry breaking. 

Turning off the $SU(2)_{L}$ triplet VEV, $v_{L}=0$, and 
assuming for simplicity that the $SU(2)_{L}\times 
SU(2)_{R}$ gauge coupling constants satisfy $g_L=g_R=g$, there are five 
fundamental parameters in the gauge/Higgs sector,
\begin{equation}
\biggl( g, \; g^\prime, \; \kappa, \; \kappa^\prime, \; v_{R} \biggr) \; .
\end{equation}
We can equivalently choose 
\begin{equation}
\biggl( \alpha, \; M_{W_{1}}^{2}, \; M_{W_{2}}^{2}, 
\; M_{Z_{1}}^{2}, \; M_{Z_{2}}^{2} \biggr) 
\end{equation}
as input parameters. The counter term for the weak mixing angle is then 
defined through these parameters and their counter terms.
Assuming that the heavy gauge bosons are much heavier 
 than the SM gauge bosons, $M_{W_2}, \, M_{Z_2} \gg M_{W_1}, \, M_{Z_1}$, 
then to leading order ${\cal O}(M_{W_1}^{2}/M_{W_2}^{2})$, 
the counterterm $\delta \hat{s}_{\theta}$ is given as 
follows~\cite{Czakon:1999ue,Czakon:1999ha,Czakon:2002wm},
\begin{eqnarray}
\frac{\delta \hat{s}_{\theta}^{2}}{\hat{s}_{\theta}^{2}} 
& = & 2 \frac{\hat{c}_{\theta}^{2}}{\hat{s}_{\theta}^{2}}
\frac{(\delta M_{Z_{1}}^{2} + \delta M_{Z_{2}}^{2}) 
- (\delta M_{W_{1}}^{2} 
+ \delta M_{W_{2}}^{2})}{(M_{Z_{1}}^{2}
+M_{Z_{2}}^{2})-(M_{W_{1}}^{2}+M_{W_{2}}^{2})} 
+{\cal {O}}\biggl(M_{W_1}^{2}/M_{W_2}^{2}\biggr)
\label{ds}\\
& \simeq & 
-2 \frac{\hat{c}_{\theta}^{2}}{\hat{s}_{\theta}^{2}} 
( \hat{c}_{\theta}^{2} - \hat{s}_{\theta}^{2})
\frac{\delta M_{W_{1}}^{2}}{M_{W_{2}}^{2} - M_{W_{1}}^{2}} + .....
\nonumber \\
& \simeq & \frac{\sqrt{2}G_{\mu}}{8\pi^{2}} \hat{c}_{\theta}^{2} 
\biggl( \frac{\hat{c}_{\theta}^{2}}{\hat{s}_{\theta}^{2}}-1 \biggr) 
\frac{M_{W_{1}}^{2}}{M_{W_{2}}^{2} - M_{W_{1}}^{2}} \cdot (3m_{t}^{2})
\; ,
\nonumber
\end{eqnarray}
where the effective weak mixing angle, $\hat{s}_{\theta}$, is defined 
as in Eq.~(\ref{swdef}). 
To go from the first to the second step in the above equation, 
we have used the following relation, 
$\left(M_{Z_{2}}^{2}+M_{Z_{1}}^{2}\right) 
- \left(M_{W_{2}}^{2}+M_{W_{1}}^{2}\right)
= \frac{g^{2}}{2\cos^{2}2\theta_{W}}v_{R}^{2}
\sim \frac{1}{\cos^{2}2\theta_{W}}\left(M_{W_{2}}^{2}-M_{W_{1}}^{2}\right)$.
In the third line  of Eq.~(\ref{ds}), we include only the leading top quark
mass dependence. 

When the limit $M_{W_{2}} \rightarrow \infty$ is taken, 
$\delta \hat{s}_{\theta}^{2}/\hat{s}_{\theta}^{2}$ approaches zero, 
and thus the SM 
result, $\delta \hat{s}_{\theta}^{2}/\hat{s}_{\theta}^{2} \sim m_{t}^{2}$ is 
{\it not} recovered, which is not what one would naively expect. 
One way to understand this  is that in the left-right model, 
four input parameters are held fixed, while in the SM three input 
parameters are fixed. There is thus no continuous limit which takes one 
from one case ($\rho \ne 1$ at tree level) to the other 
($\rho = 1$ at tree level). This discontinuity, which has been pointed 
out previously\cite{Lynn:1990zk,Passarino:1990xx}, is closely tied 
to the fact that the triplet Higgs boson is 
non-decoupling\cite{Toussaint:1978zm,Senjanovic:1978ee,Chen:2003fm}. 
Due to this non-decoupling effect, even if the triplet VEV is extremely small, 
as long as it is non-vanishing, there is the need for the fourth 
input parameter. The only exception to this is if there is a custodial 
symmetry which forces $v^\prime=0$: in this case only the usual 
three input parameters are necessary.

We also note that the contribution of the lightest neutral Higgs 
in this case is given by\cite{Czakon:1999ue}, 
\begin{equation}
(\Delta r)^{\mbox{\small lightest Higgs}}_{LR} = 
\frac{\sqrt{2}G_{\mu}}{48\pi^2}
\biggl(
\frac{M_{W_1}^{2}}{M_{W_2}^2} \frac{\hat{c}_{\theta}^2}{\hat{s}_{\theta}^{2}}
\bigl(1-2\hat{s}_{\theta}^{2}\bigr)
+ \frac{M_{W_1}^{2}}{M_{Z_2}^2} \frac{1}{\hat{s}_{\theta}^{2}}
\bigl(4\hat{c}_{\theta}^{2}-1\bigr)
\biggr) M_{H^0}^{2} \; ,
\end{equation}
which depends on $M_{H^0}$ quadratically, and is suppressed 
by the heavy gauge boson masses, 
$M_{W_{2}}^{2}$ and $M_{Z_{2}}^{2}$.  
The contributions of the remaining scalar fields also 
have a similar structure.

\section{Minimization of the Scalar Potential in Model with a Triplet}
\label{minimize}

In this section, we summarize our results on minimization of the scalar potential 
in the model with a triplet Higgs. 
From the minimization conditions, 
$\frac{\partial V}{\partial \eta^{0}} 
\bigg|_{<H>,<\Phi>} = 
\frac{\partial V}{\partial \phi^{0}}
\bigg|_{<H>,<\Phi>} =0$,
we obtain the following conditions,
\begin{eqnarray}
4\mu_{2}^{2} t_{\delta} + \lambda_{2}v^{2}t_{\delta}^{3}
+2\lambda_{3}v^{2}t_{\delta}-4\lambda_{4}v & = & 0
\\
\mu_{1}^{2} + \lambda_{1}v^{2} + \frac{1}{8} 
\lambda_{3}v^{2}t_{\delta}^{2} 
-\frac{1}{2}\lambda_{4}vt_{\delta} & = & 0
\; .
\end{eqnarray}
We ues the short hand notaion, $t_{\delta}=\tan\delta$. 
The second derivatives are,
\begin{eqnarray}
\frac{\partial^{2}V}{\partial \eta^{+} \partial \eta^{-}}
\bigg|_{<H>,<\Phi>}
& = & \mu_{2}^{2} + \frac{1}{8}\lambda_{2}v^{2}t_{\delta}^{2}
+\frac{1}{2}\lambda_{3} v^{2}
\\
\frac{\partial^{2}V}{\partial \eta^{+} \partial \phi^{-}}
\bigg|_{<H>,<\Phi>} & = &
\frac{\partial^{2}V}{\partial \eta^{-} \partial \phi^{+}}
\bigg|_{<H>,<\Phi>}
 =  \lambda_{4} v
\label{chargemix2}\\
\frac{\partial^{2}V}{\partial \phi^{+} \partial \phi^{-}}
\bigg|_{<H>,<\Phi>}
& = & 
\mu_{1}^{2}+\lambda_{1}v^{2}
+\frac{1}{8}\lambda_{3}v^{2}t_{\delta}^{2}
+\frac{1}{2} \lambda_{4} v t_{\delta}
\\
\frac{\partial^{2}V}{\partial \eta^{0} \partial \eta^{0}}
\bigg|_{<H>,<\Phi>}
&=&
\mu_{2}^{2}+\frac{3}{4}\lambda_{2}v^{2}t_{\delta}^{2}
+\frac{1}{2} \lambda_{3}v^{2}
\\
\frac{\partial^{2}V}{\partial \eta^{0} \partial \phi^{0}}
\bigg|_{<H>,<\Phi>} &=&
\frac{1}{2} \lambda_{3} v^{2} t_{\delta}
-\lambda_{4}v
\label{neutralmix1}
\\
\frac{\partial^{2}V}{\partial \phi^{0}\partial \phi^{0}}
\bigg|_{<H>,<\Phi>}
&=&
\mu_{1}^{2}+3\lambda_{1}v^{2}+\frac{1}{8}\lambda_{3}v^{2}t_{\delta}^{2}
-\frac{1}{2}\lambda_{4}vt_{\delta} \; .
\end{eqnarray}
If $\frac{\partial^{2}V}{\partial \eta^{+} \partial \phi^{-}}=0$, 
then there is no mixing between the doublet and the triplet. 
This requires $\lambda_{4} =0$.



\begin{thebibliography}{0}

\bibitem{ewpewwg}
{LEP Electroweak Working Group, http://lepewwg.web.cern.ch/LEPEWWG.}

\bibitem{Peskin:1991sw}
  M.~ E.~Peskin and T.~ Takeuchi,
  {\sl Phys. Rev.} {\bf D46}, 381 (1992). 

\bibitem{altarelli:1991a}
  G. Altarelli and R. Barbieri, 
  {\sl Phys. Lett.} {\bf B253}, 161 (1991).

\bibitem{pdg04}
S. Eidelman {\it et al} (PDG 2004), 
\emph{Review Of Particle Physics. Particle Data Group}, 
{\sl Phys. Lett.} {\bf B592}, 1 (2004).

\bibitem{Passarino:1990xx}
G.~Passarino,
{\sl Nucl. Phys.} \textbf{B361},
351 (1991).


\bibitem{Lynn:1990zk}
B.~W. Lynn and E.~Nardi,
{\sl Nucl. Phys.} \textbf{B381},
467 (1992).

\bibitem{chivukula:1999}
S. Chivukula and N. Evans, 
{\sl Phys. Lett.} {\bf B464}, 244 (1999).

\bibitem{Chivukula:2000px}
 R.~S.~Chivukula, C.~Hoelbling and N.~J.~Evans,
  {\sl Phys. Rev. Lett.}  {\bf 85}, 511 (2000).  


\bibitem{Peskin:2001rw}
  M.~E.~Peskin and J.~D.~Wells,
 {\sl Phys. Rev.}  {\bf D64}, 093003 (2001).

\bibitem{Chen:2003fm}
  M.-C.~Chen and S.~Dawson,
{\sl  Phys. Rev.} {\bf D70}, 015003 (2004).


\bibitem{Chen:2004ig}
  M.-C.~Chen and S.~Dawson,
in {\sl Tsukuba 2004, Supersymmetry and unification of fundamental 
  interactions, 921-924} [hep-ph/0409163].


\bibitem{Marciano:1980pb}
 A.~Sirlin,
 {\sl Phys. Rev.} {\bf D22}, 971 (1980). 

\bibitem{marciano:1980a}
  W.~J.~Marciano and A.~Sirlin,
{\sl  Phys. Rev.} {\bf D22}, 2695 (1980)
  [Erratum-ibid. {\bf D31}, 213 (1985)]. 

\bibitem{Sirlin:1981yz}
  A.~Sirlin and W.~J.~Marciano,
{\sl  Nucl. Phys.} {\bf B189}, 442 (1981).



\bibitem{jegerlehner}
F. Jegerlehner, 
\emph{Renormalizaing the Standard Model},   
lecture given at Theoretical Advanced Study Institute in 
Elementary Particle Physics (TASI 90), Boulder, 
CO, June 3-29, 1990. Published in {\sl Boulder TASI 90:476-590}.


\bibitem{Erler:2004nh}
  J.~Erler and P.~Langacker,
  \emph{Electroweak model and constraints on new physics},
  published in Ref.~\refcite{pdg04} [hep-ph/0407097].


\bibitem{pierce}
D.M. Pierce, 
\emph{Renormalization of Supersymmetric Theories}, 
lecture given at the Theoretical Advanced Study Institute in 
Elementary Particle Physics (TASI 97), Boulder, CO, June 1-7, 1997.
Published in {\sl Boulder 1997: Supersymmetry, Supergravity 
and Supercolliders}, 343-389 [hep-ph/9805497].


\bibitem{Hollik:1993cg}
  W.~Hollik,
MPI-PH-93-21. 

\bibitem{degrassi:1992a}
G. DeGrassi and A. Sirlin, 
{\sl Nucl. Phys.} {\bf B383}, 73 (1992).


\bibitem{Chanowitz:1978uj}
 M.~J.~G.~Veltman,
  {\sl Nucl. Phys.} {\bf B123}, 89 (1977). 

\bibitem{chanowitz:1978a}
  M.~S.~Chanowitz, M.~A.~Furman and I.~Hinchliffe,
  {\sl Phys. Lett.} {\bf B78}, 285 (1978).



\bibitem{Marciano:2004hb}
W.~J. Marciano, 
hep-ph/0411179.


\bibitem{Logan:1999if}
  H.~E.~Logan, Ph.D thesis [hep-ph/9906332]. 

\bibitem{logan:2000a}
  H.~E.~Haber and H.~E.~Logan,
{\sl  Phys. Rev.} {\bf D62}, 015011 (2000). 


\bibitem{Blank:1997qa}
T.~Blank and W.~Hollik,
{\sl Nucl. Phys.} \textbf{B514},
113 (1998) [hep-ph/9703392].


\bibitem{Consoli:1989pc}
  M.~Consoli, W.~Hollik and F.~Jegerlehner,
CERN-TH-5527-89.



\bibitem{Toussaint:1978zm}
D.~Toussaint,
{\sl Phys. Rev.} \textbf{D18}, 1626 (1978).


\bibitem{Senjanovic:1978ee}
G.~Senjanovic and A.~Sokorac,
{\sl Phys. Rev.} \textbf{D18},
2708 (1978).


\bibitem{Forshaw:2003kh}
  J.~R.~Forshaw, A.~Sabio Vera and B.~E.~White,
{\sl  JHEP} {\bf 0306}, 059 (2003)  
 [arXiv:hep-ph/0302256].

\bibitem{Passarino:1990nu}
  G.~Passarino,
  {\sl Phys. Lett.} {\bf B247}, 587 (1990). 

\bibitem{Gunion:1990dt}
  J.~F.~Gunion, R.~Vega and J.~Wudka,
 {\sl Phys. Rev.} {\bf D43}, 2322 (1991). 



\bibitem{Pomarol:1993mu}
  A.~Pomarol and R.~Vega,
 {\sl Nucl. Phys.} {\bf B413}, 3 (1994). 

\bibitem{Forshaw:2001xq}
  J.~R.~Forshaw, D.~A.~Ross and B.~E.~White,
  {\sl JHEP} {\bf 0110}, 007 (2001). 


\bibitem{Czakon:1999ue}
M.~Czakon, M.~Zralek, and J.~Gluza, 
{\sl  Nucl. Phys.} \textbf{B573}, 57 (2000)  
 [hep-ph/9906356].


\bibitem{Czakon:1999ha}
M.~Czakon, J.~Gluza, F.~Jegerlehner and M.~Zralek,
{\sl  Eur. Phys. J.} \textbf{C13}, 275 (2000) 
[hep-ph/9909242].


\bibitem{Czakon:2002wm}
  M.~Czakon, J.~Gluza and J.~Hejczyk,
  {\sl Nucl. Phys.} {\bf B642}, 157 (2002). 




\bibitem{Csaki:2002qg}
  C.~Csaki, J.~Hubisz, G.~D.~Kribs, P.~Meade and J.~Terning,
  {\sl Phys. Rev.} {\bf D67}, 115002 (2003). 


\bibitem{Hewett:2002px}
  J.~L.~Hewett, F.~J.~Petriello and T.~G.~Rizzo,
  {\sl JHEP} {\bf 0310}, 062 (2003). 


\bibitem{Veltman:1977kh}
M.~J.~G.~Veltman,
{\sl Nucl.\ Phys.} {\bf B123}, 89 (1977).


\bibitem{Pati:1974yy}
  J.~C.~Pati and A.~Salam,
{\sl  Phys. Rev.} {\bf D10}, 275 (1974). 

\bibitem{Mohapatra:1974hk}
  R.~N.~Mohapatra and J.~C.~Pati,
  {\sl Phys. Rev.} {\bf D11}, 566 (1975). 

\bibitem{Mohapatra:1974gc}
  R.~N.~Mohapatra and J.~C.~Pati,
  {\sl Phys. Rev.} {\bf D11}, 2558 (1975). 

\bibitem{Senjanovic:1975rk}
  G.~Senjanovic and R.~N.~Mohapatra,
  {\sl Phys. Rev.} {\bf D12}, 1502 (1975).



\bibitem{Chen:2004ww}
M.-C. Chen and K.~T. Mahanthappa,
 {\sl Phys. Rev.} \textbf{D71},
  035001 (2005). 

\bibitem{chen:2003a}
M.-C. Chen and K.~T. Mahanthappa, 
{\sl Int. J. Mod. Phys.} {\bf A18}, 5819 (2003). 

\bibitem{chen:2004a}
M.-C. Chen and K.~T. Mahanthappa,
{\sl AIP Conf. Proc.}  {\bf 721}, 269 (2004). 

\bibitem{chen:2000a}
M.-C. Chen and K.~T. Mahanthappa,
{\sl Phys. Rev.} {\bf D62}, 113007 (2000). 

\bibitem{chen:2001a}
M.-C. Chen and K.~T. Mahanthappa,
{\sl Phys. Rev.}  {\bf D65}, 053010 (2002). 

\bibitem{chen:2002a}
M.-C. Chen and K.~T. Mahanthappa,
{\sl Phys. Rev.} {\bf D68}, 017301 (2003). 

\bibitem{chen:2004b}
M.-C. Chen and K.~T. Mahanthappa,
{\sl Phys. Rev.} {\bf D70}, 113013 (2004). 


\bibitem{chen:2004c}
M.-C. Chen and K.~T. Mahanthappa,
in {\sl Tsukuba 2004, SUSY 2004, 761-764} [hep-ph/0409165]. 

\bibitem{chen:2005a}
M.-C. Chen, 
{\sl Phys.\ Rev.}  {\bf D71}, 113010 (2005). 



\end{thebibliography}
\end{document}